\RequirePackage{fix-cm}
\documentclass[twocolumn]{svjour3}          
\smartqed  
\usepackage{graphicx}
\usepackage{amsmath}    
\usepackage{epstopdf}   
\def\la{\mathrel{\mathchoice {\vcenter{\offinterlineskip\halign{\hfil
$\displaystyle##$\hfil\cr<\cr\sim\cr}}}
{\vcenter{\offinterlineskip\halign{\hfil$\textstyle##$\hfil\cr
<\cr\sim\cr}}}
{\vcenter{\offinterlineskip\halign{\hfil$\scriptstyle##$\hfil\cr
<\cr\sim\cr}}}
{\vcenter{\offinterlineskip\halign{\hfil$\scriptscriptstyle##$\hfil\cr
<\cr\sim\cr}}}}}
\def\ga{\mathrel{\mathchoice {\vcenter{\offinterlineskip\halign{\hfil
$\displaystyle##$\hfil\cr>\cr\sim\cr}}}
{\vcenter{\offinterlineskip\halign{\hfil$\textstyle##$\hfil\cr
>\cr\sim\cr}}}
{\vcenter{\offinterlineskip\halign{\hfil$\scriptstyle##$\hfil\cr
>\cr\sim\cr}}}
{\vcenter{\offinterlineskip\halign{\hfil$\scriptscriptstyle##$\hfil\cr
>\cr\sim\cr}}}}}
\def\deg{\hbox{$^\circ$}}

\newcommand{\beq}{\begin{equation}}
\newcommand{\eeq}{\end{equation}}
\newcommand{\bdi}{\begin{displaymath}}
\newcommand{\edi}{\end{displaymath}}

\begin{document}

\title{Laboratory measurements of super-resolving Toraldo pupils for 
radio astronomical applications   
%
}

\titlerunning{Laboratory measurements of Toraldo pupils}        

\author{Luca Olmi \and
        Pietro Bolli \and
	Luca Cresci \and
	Francesco D'Agostino \and
	Massimo Migliozzi \and
	Daniela Mugnai \and
	Enzo Natale \and
	Renzo Nesti \and
	Dario Panella \and
	Lorenzo Stefani
}


\institute{L. Olmi \at
	      Istituto Nazionale di Astrofisica (INAF), 
	      Osservatorio Astrofisico di Arcetri, Largo E. Fermi 5,
              I-50125 Firenze, Italy \\
	      \email{olmi.luca@gmail.com} \\
	      and \\	
	      University of Puerto Rico, Rio Piedras Campus, Physics Dept., Box 23343,
              UPR station, San Juan, Puerto Rico, USA \\
        \and
           P. Bolli \at
              Istituto Nazionale di Astrofisica (INAF),
              Osservatorio Astrofisico di Arcetri, Largo E. Fermi 5,
              I-50125 Firenze, Italy \\
        \and
           L. Cresci \at
              Istituto Nazionale di Astrofisica (INAF),
              Osservatorio Astrofisico di Arcetri, Largo E. Fermi 5,
              I-50125 Firenze, Italy \\
        \and
           Francesco D'Agostino \at 
	   Universit\`a di Salerno, Via G. Paolo II 132, I-84084 Fisciano (SA), Italy \\
        \and
        Massimo Migliozzi  \at
	   Universit\`a di Salerno, Via G. Paolo II 132, I-84084 Fisciano (SA), Italy \\
        \and
        Daniela Mugnai \at 
	   Consiglio Nazionale delle Ricerche (CNR), Istituto di Fisica Applicata Nello Carrara,
	   Via Madonna del Piano 10, I-50019 Firenze, Italy \\
        \and
        Enzo Natale \at 
              Istituto Nazionale di Astrofisica (INAF),
              Osservatorio Astrofisico di Arcetri, Largo E. Fermi 5,
              I-50125 Firenze, Italy \\
 	\and
        Renzo Nesti \at
              Istituto Nazionale di Astrofisica (INAF),
              Osservatorio Astrofisico di Arcetri, Largo E. Fermi 5,
              I-50125 Firenze, Italy \\
        \and
        Dario Panella \at
              Istituto Nazionale di Astrofisica (INAF),
              Osservatorio Astrofisico di Arcetri, Largo E. Fermi 5,
              I-50125 Firenze, Italy \\
        Lorenzo Stefani \at
           Consiglio Nazionale delle Ricerche (CNR), Istituto di Fisica Applicata Nello Carrara,
           Via Madonna del Piano 10, I-50019 Firenze, Italy \\
}


\date{Received: date / Accepted: date}

\maketitle

\begin{abstract}
The concept of super-resolution refers to various methods for improving the angular
resolution of an optical imaging system beyond the classical diffraction limit.
Although several techniques to narrow the central lobe of the illumination Point Spread Function
have been developed in optical microscopy, most of these methods cannot be implemented 
on astronomical telescopes. A possible exception is represented by the variable transmittance 
filters, also known as  ``Toraldo Pupils" (TPs) since they were  introduced  for the first time 
by G. Toraldo di Francia in 1952~\cite{toraldo1952a}. In the microwave range, 
the first successful laboratory test of TPs was performed in 2003~\cite{mugnai2003}. 
These first results suggested that TPs could represent a viable approach to achieve 
super-resolution in Radio Astronomy.
We have therefore started a 
project devoted to a more exhaustive analysis of TPs, in order to assess their potential 
usefulness to achieve super-resolution on a 
radio telescope, as well as to determine their drawbacks.  
In the present work we report on the results of extensive
microwave measurements, using TPs with different geometrical shapes,
which confirm the correctness of the first experiments in 2003. We have also extended
the original investigation to carry out full-wave electromagnetic numerical 
simulations and also to perform
planar scanning of the near-field and transform the results into the far-field.
\keywords{Angular resolution \and super-resolution techniques \and Toraldo pupils 
\and microwave measurements \and near-field}
%
\end{abstract}

\section{Introduction}
\label{sect:intro}  

The concept of super-resolution refers to various methods for improving the angular
resolution of an optical imaging system beyond the classical diffraction limit.
In optical microscopy, several techniques have been successfully developed 
with the aim of narrowing the central lobe of the illumination Point Spread Function (PSF).
These techniques either involve changing the fluorescence status of the specimen, or 
the specimen is imaged within a region having a radius much shorter than the illumination 
wavelength, thus exploiting the unique properties of the evanescent waves. 
%
Using electrically small artificial structures, negative refractive index (NRI) metamaterials 
(also known as negative index media, or NIM) provide a physical 
platform to controlling the properties of electromagnetic (EM) waves.
One of the most striking properties of NRI materials is that a slab of metamaterial 
can be a  ``perfect lens'' in which the evanescent waves, instead of decaying, are in fact enhanced through the slab
and in theory it is thus capable of imaging infinite small features of targets~\cite{zhang2008}. 

However, few efforts have been made to overcome the diffraction limit of telescopes. 
This is mainly attributed to the fact that remote objects, like astronomical targets, are not easily accessible 
for artificial radiation manipulation and the great size of telescopes reduces the possibility of super-resolution 
optical elements composed of metamaterials. A concept that used a NRI lens positioned between the
conventional reflector of a radio telescope and its focal plane  to shape the PSF has been proposed~\cite{may2004}
but, to our knowledge, there are no published experimental measurements.  A more exotic concept based on
quantum cloning~\cite{kellerer2014} has also been published, but its practical realization is yet to be proven.

In a classical filled-aperture telescope with diameter $D$, and angular resolution $\simeq \lambda/D$, the purpose 
of a super-resolving optical device would be to increase the resolving power of the telescope without 
increasing its aperture. In fact, aperture synthesis telescopes can enhance the angular resolution beyond the 
limits of its individual filled-aperture telescopes, but at the cost of much increased complexity.
References~\cite{cagigal2004} and \cite{canales2004} review and discuss  methods for designing 
super-resolving pupil masks that use {\it variable transmittance pupils} 
for optical telescopes. These pupils are attractive to design antennas and telescopes with 
resolution significantly better than the diffraction limit, $\simeq \lambda/D$, 
since their realization does not require significant modifications to the optical layout of the telescope
or any new technological breakthrough.

The first time such pupils were discussed was at a lecture delivered by 
Toraldo di Francia at a colloquium on optics and microwaves in 1952~\cite{toraldo1952a}.
Toraldo di Francia suggested that the classical limit of optical resolution could be improved 
interposing a filter consisting of either infinitely narrow concentric rings or finite-width 
concentric annuli of different amplitude and phase transmittance in the entrance pupil of an optical system.
These pupils are now also known as Toraldo pupils (TPs, hereafter) and are considered a special 
case  of a the more general case of variable transmittance pupils. In fact, it can be easily shown that 
a TP consisting of infinitely narrow concentric rings, or {\it continuous} Toraldo pupil, is equivalent
to a transmittance pupil with a complex illumination function (Olmi {\it et al.}, in prep.).
Many other super-resolving filters have since been proposed, but these methods offer little 
theoretical advantage over the original method proposed by Toraldo di Francia, as shown in Ref.~\cite{cox1982}.

TPs have been widely analyzed in the context of microscopy~\cite{neil2000,martinez2002,kim2012}, but so far they 
have never been designed for telescope or antenna applications. In fact, the first experimental studies in the 
microwave range of a TP were carried out in 2003~\cite{mugnai2003} and 2004~\cite{ranfagni2004}. 
These successful laboratory results later raised the interest in the potential application of TPs to microwave antennas and 
radio telescopes. 
In fact, given that discrete TPs (i.e., employing finite-width concentric coronae with different complex transmittance) 
for the microwave range are very easy to fabricate and relatively easy to model, 
we started a project devoted to a more careful analysis of TPs and how they could be
implemented on a radio telescope. 

During the first part of this work we have conducted extensive electromagnetic numerical     
simulations of TPs, using a commercial full-wave software tool, that have already been 
discussed elsewhere~\cite{olmi2016}. 
We have used these simulations to study various EM effects that can mask and/or modify the 
performance of the pupils and to analyze the near-field (NF) as well as the far-field (FF) response.  
We then used these EM  simulations to prepare more comprehensive laboratory testing, 
and the purpose of this paper is thus to discuss a series of experimental tests conducted at 20\,GHz that 
significantly extend and improve previous laboratory investigations~\cite{mugnai2003,ranfagni2004}. Our results 
again confirm Toraldo di Francia's model and also 
suggest that TPs should be investigated as a potential tool to achieve super-resolution on a radio telescope.

The outline of the paper is as follows. In Sect.~\ref{sec:toraldo} we give an overview of the basic properties of TPs. 
In  Sect.~\ref{sec:EM} we summarize the EM simulations which have been described elsewhere~\cite{olmi2016} and 
also discuss some additional EM modeling specific to our laboratory tests.  In Section~\ref{sec:exp}
we describe the laboratory setup and procedures used to perform our measurements and the results obtained 
in the NF using two different types of TP. Therefore, in Sect.~\ref{sec:nf2ff} we describe how the NF measurements
were converted into the FF. Finally, in Sect.~\ref{sec:concl} we draw our conclusions.

\section{TORALDO PUPILS}
\label{sec:toraldo}

\subsection{Analytical description}
\label{sec:analytical}

As we earlier mentioned, Toraldo di Francia introduced for the first time the concept of 
variable transmittance pupils in 1952~\cite{toraldo1952a,toraldo1952b},  and an analytical description of 
TPs can be found in several references~\cite{born1999,cox1982,mugnai2003}. 
For the convenience of the reader we briefly review the theory here.
In Toraldo di Francia's model several approximations are implicitly assumed, thus reducing the problem 
of the scattered fields to a scalar diffraction problem in the case of linearly polarized radiation 
incident on the aperture (see, e.g., Chapter 5.14 of Ref.~\cite{silver}). 
A full-wave discussion of the scattered fields from a TP is beyond the scopes of this work, and therefore 
we will adopt the same approximations as in Toraldo di Francia's original version. However,
for the near- to far-field transformation described in Section~\ref{sec:nf2ff} we will 
adopt the full-wave approach both in our laboratory measurements and in the EM simulations.

Let us then consider a circular pupil of diameter $D$ and divide it into $n$ discrete, concentric
circular coronae by means of $n + 1$ circumferences with diameters $\alpha_0 D$, $\alpha_1 D$, 
$...$, $\alpha_n D$, where $\alpha_0$, $...$, $\alpha_n$ is a succession of numbers in increasing 
order, with $\alpha_0=0$ and $\alpha_n=1$. In Toraldo di Francia's original version, each corona
is either perfectly transparent or provides a phase inversion (i.e., a $\Delta \phi = 180^{\circ}$ 
phase change, see below), and is illuminated by a plane wave (i.e., with uniform phase over the pupil).
By setting $x = \pi \frac{D}{\lambda} \sin \theta$, 
where $\theta$ is the angle of diffraction measured with respect to the optical axis and $\lambda$ is the wavelength, 
it can be shown that the total amplitude  (in the FF), $A(x)$,  diffracted by the composite TP is given by:
\begin{equation}
A(x) = \sum_{i=0}^{n-1} \frac{k_{i+1}}{x} [ \alpha_{i+1} J_1(\alpha_{i+1} x) - \alpha_{i} J_1(\alpha_{i} x) ] 
\label{eq:ampl}
\end {equation}
where $k_{i+1} = \frac{\pi D^2 } {2 \lambda^2 } A_{i+1}$ is a constant that is proportional 
to the amplitude $A_{i+1}$ illuminating each corona, and $J_1$ is the Bessel function of the first order. 

Once the partition of the pupil into $n$  circular coronae is established, we can 
impose $n$ independent conditions on $A(x)$, thus obtaining a 
system of $n$ equations from which we can determine the 
coefficients $k_1,...,k_n$. These equations can be obtained, for example, by $A(x_i) = 0$ 
where $x_i$ ($i=0...n-1$) represent the zeros of the diffracted amplitude. In particular, by
setting the position of the first zero the user establishes the width of the main lobe and thus 
the extent of the super-resolution effect.

Note that the $k$ coefficients can also have negative values, 
in which case they represent a {\it phase-inversion} of the wave propagating through the pupil. 
In fact, in this work we are only concerned with the simplest possible pupil, i.e., composed
of discrete circular coronae that do or do not introduce a phase-inversion. 
A description of continuous phase filters can be found elsewhere~\cite{dejuana2003}, 
as well as their application to ground-based telescopes~\cite{cagigal2004}, while the 
precise relation with discrete Toraldo pupils will be discussed in a forthcoming paper (Olmi {\it et al.}, in prep.) .
Note that in order to obtain the phase-inversion, $\Delta \phi = (2\pi/\lambda)\, \Delta l = \pi$, the
optical path {\it excess} (with respect to the wave propagating through air) must be
$\Delta l =  l_{\rm diel} - l_{\rm air}  = ( n_{\rm diel} - n_{\rm air}  ) \, \Delta s = 0.5 \lambda$,
where $\Delta s$ is the physical thickness of the dielectric material and $n$ is the
refraction index.  If $n_{\rm diel} \simeq 1.5$ and $n_{\rm air} \simeq 1$ then $\Delta s \simeq \lambda$.

\subsection{Illumination of the pupil}
\label{sec:illum}

Solving the set of equations $A(x_i) = 0$ allows to determine the $k_i$ coefficients and thus,
given that they are  proportional  to the amplitude $A_i$ illuminating each corona, it also
determines the type of amplitude {\it apodization} required by a specific TP. In general, if no {\it a priori} 
constraint on the illumination is set, the resulting apodization may be quite complex, and while at visible 
wavelengths the required amplitude distribution could in principle be obtained using the appropriate 
neutral filters, in the microwave range this technique is not easily implemented. For example, 
in Ref.~\cite{olmi2016} it is shown that for the simple case of a 3-coronae TP, its geometry 
can be chosen so that the required intensity illumination at the center of each corona 
can be adequately fit by a Gaussian beam. At microwave wavelengths a Gaussian illumination 
can be easily obtained using, for example, a rectangular feedhorn but a Gaussian beam  cannot 
provide a uniform phase over the pupil, which is one of the optical conditions required by a TP. The pupil could be located in the 
FF of the Gaussian beam source, where the spherical wavefront can be approximated as a plane
wave over the extent of the pupil, thus satisfying the uniform phase condition but without the
required amplitude apodization.

   \begin{figure*}  
   \begin{center}
   \begin{tabular}{c} 
   \includegraphics[totalheight=8cm]{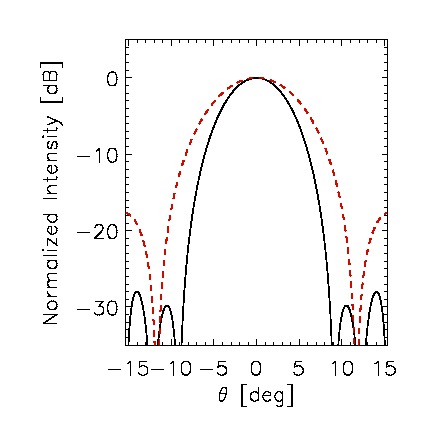}   
   \includegraphics[totalheight=8cm]{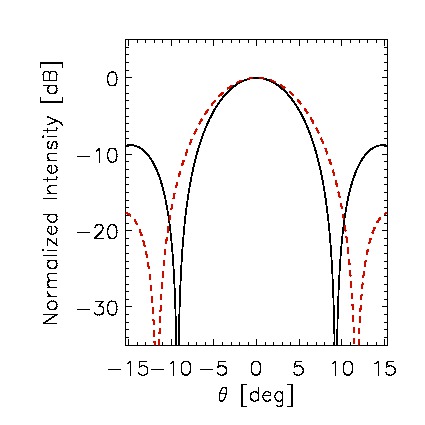} 
   \end{tabular}
   \end{center}
   \caption[example]
   { \label{fig:illum4cor}
{\it Left panel.}
Diffraction pattern at $\nu = 20\,$GHz by a four-coronae TP, as given by the square 
of Eq.\,(\ref{eq:ampl}) for $n = 4$ (black solid line), and that of a normal pupil of 
equal diameter (red dashed line). Both curves are normalized with respect to the on-axis value. 
The correct amplitude apodization has been applied resulting in an angular resolution gain $G=1.94$.
{\it Right panel.}
Normalized diffraction pattern at $\nu = 20\,$GHz 
assuming the correct phase distribution but with no amplitude apodization, resulting in $G=1.65$.
}
   \end{figure*}

Since it appears very difficult to implement amplitude apodization and uniform phase simultaneously 
over the pupil it is of interest to analyze how the output amplitude from a TP would change if one
of these conditions is not met. For example, if the phase inversion, corresponding to negative 
$k$ coefficients, is {\it not} applied then no super-resolution effect is achieved and the shape of the 
PSF is basically unchanged. Instead, the condition about the amplitude apodization can be relaxed,
though at the cost of a mild reduction in the super-resolution effect and modified sidelobes. 

The simple case of a 3-coronae TP
has been discussed elsewhere~\cite{olmi2016}, and we thus show the equivalent effects for the
case of a 4-coronae TP in Fig.~\ref{fig:illum4cor}. For this specific example we have selected
the $\alpha$ coefficients as $\alpha = [0,0.16,0.4,0.8,1]$. Given that an open pupil with the 
same diameter has its first zero at $x=3.83$ (corresponding to $\sin \theta =1.22 \lambda/D$) 
we set the amplitude $A(x_i) = 0$ at the values of $x=[3,4,5]$, thus narrowing the main lobe, and
we also set the maximum value of $A$ at $x=0$. 
The resulting (unnormalized)  $k$ coefficients are $k = [-778.2, 264.5, -40.8, 22.9]$.
The left panel of Fig.~\ref{fig:illum4cor} shows the calculated diffracted amplitude by the TP assuming
the correct amplitude apodization and phase inversion. It is customary~\cite{cox1982}  to define the 
resolution gain\footnote{Not to be confused with the {\it antenna} gain.}, $G$,
as the ratio of the radius of the first zero of the Airy distribution 
to that of the super-resolving diffraction pattern. Here we replace the radius of the first zero with the
full width at half-maximum (FWHM, corresponding to the width at $-3\,$dB with respect to the peak value) 
of the PSF to measure $G$, which in this case is  $G=1.94$. 
In the right panel of Fig.~\ref{fig:illum4cor} we show the diffracted amplitude with {\it no} amplitude apodization applied, but still
with the correct phase relations. The relative intensity, number and position of the sidelobes have changed 
and the super-resolution effect is somewhat lower, with $G=1.65$. 


This example shows that, although the super-resolution effect is still present, the performance of a TP is different if
the correct apodization condition is not applied, which makes these pupils difficult to study in the 
microwave range with EM numerical simulations (see Sect.~\ref{sec:EM}). Likewise, testing TPs
in the laboratory would in principle require to design a suitable method to implement  both
amplitude apodization and phase inversions. However, in Sect.~\ref{sec:exp} we show that 
successful  results can also be obtained when amplitude apodization is not present. These problems
could be partly resolved with the use of phase-only masks~\cite{dejuana2003,liu2008} which will also be 
discussed in a future work (Olmi {\it et al.}, in prep.).

\section{ELECTROMAGNETIC SIMULATIONS}
\label{sec:EM}

\subsection{Summary of previous results}
\label{sec:old}

Any measurement performed in the laboratory cannot fully satisfy all of the optically ideal
conditions which are assumed for a TP, and thus a way to validate and test the performance 
of a TP under less than ideal laboratory conditions is required. The method used must also 
be able to address how the theoretical performance of a TP can be masked or altered by 
various optical (mostly diffractive) effects. Therefore, we carried out an extensive series of
EM numerical simulations using the commercial software  
FEKO\footnote{http://www.altairhyperworks.com/product/FEKO},
a comprehensive EM simulation software tool for the electromagnetic field 
analysis of 3D structures. 

After a few attempts with frequencies, $\nu$, of 10 and 50\,GHz, we decided to perform the bulk of 
the EM simulations, followed by laboratory measurements, at $\nu=20\,$GHz. At lower frequencies
the computing time would be much lower, but the size (and separations) of the optical elements in units of
wavelengths would be uncomfortably large for our expected experimental setup. At higher frequencies
the situation pretty much reverses, with far too long computing times though with the advantage of 
reasonably sized microwave components. We found that 20\,GHz represents a good trade-off between 
these opposite requirements.

During the first part of this work our EM simulations had the main purpose of generating a FEKO 
model which would represent an optical configuration as close as possible to the ideal system 
described in Sect.~\ref{sec:toraldo}, thus trying to simulate the performance predicted by Eq.~(\ref{eq:ampl}). 
These numerical simulations have already been discussed in a previous work~\cite{olmi2016}, 
and thus here we just present a brief summary of the main results:
\begin{itemize}
\item[1.] Two different sources have been tested: plane-wave and Gaussian-beam
illumination. The use of plane-waves requires a special technique to simulate an infinite
ground plane where the pupil is realized, in order to avoid strong diffraction effects from the edges 
of a finite surface.
\item[2.] The Gaussian beam produced by a rectangular feedhorn must illuminate the pupil
in the FF to ensure the uniform phase condition. The tapered illumination of the feedhorn
can also be used with a finite ground screen.
\item[3.] Both NF and FF distributions have been analyzed. However, the analytical
model of a TP discussed in Sect.~\ref{sec:analytical} can only be compared with the FF 
numerical results. 
\item[4.] The simulated FFs confirm the super-resolution effect, even with no amplitude apodization.
They also show that TPs with different numbers of coronae can be used to achieve a trade-off between
resolution gain, $G$, strength and position of the side lobes, and overall efficiency, measured as the 
decrease of the on-axis intensity compared to the open pupil.
\end{itemize}
Our initial EM simulations were thus successful, and helped us to analyze  both the super-resolution
effect using various geometrical configurations, and also other diffraction effects that could mask the
expected narrowing of the PSF. Although simulating the FF is in general easier and less time-consuming 
than computing the NF, we were expecting to perform our experimental measurements in the NF 
(see Sect.~\ref{sec:prelim}). Therefore, we also carried out a number of EM simulations of the NF that 
were more closely reproducing the laboratory conditions, which we describe in the next section.

   \begin{figure}  
   \begin{center}
   \begin{tabular}{c} 
   \includegraphics[height=8.0cm]{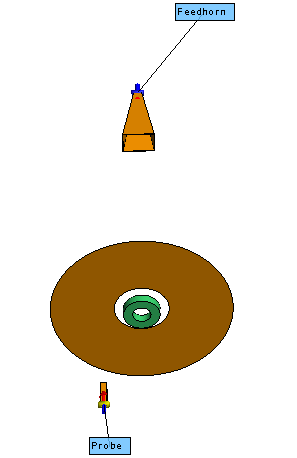}     
   \end{tabular}
   \end{center}
   \caption[example]
   { \label{fig:sparam}
FEKO model used for the numerical simulations of the $S$-parameters. The three-coronae
TP is shown here, with the on-axis feedhorn and the probe positioned in an off-axis position after the pupil.
}
   \end{figure}

\subsection{Numerical simulations for S-parameters}
\label{sec:spar}

One of the main goals of this work was to compare our experimental measurements in the NF with the 
expected results from our EM simulations. However, in this respect the main limitation of the 
EM simulations summarized above and discussed in Ref.~\cite{olmi2016} is that they used 
an incident plane wave as the illuminating source. Furthermore, these simulations sampled 
the NF point-by-point, without taking into account the finite spatial response of the NF probe
used in the actual measurements. 
Hence, the only viable option to perform a more
reliable comparison between the measured and simulated diffraction patterns in the NF
was to include the NF probe as well as the feedhorn in the FEKO model.

Therefore, we built a FEKO model specific for the NF measurements, that included both the feedhorn as the launcher and the probe
after the pupil to sample the NF. Figure~\ref{fig:sparam} shows our model where the feedhorn is visible behind the three-coronae pupil and the 
NF probe can be seen in front of the screen, in an off-axis position. In this case, since the field detection is done only through
the excitation of the waveguide, the EM simulation returns the values of the scattering parameters, or $S$-parameters,
which measure the reflection and transmission coefficients of a 2-port device, such as a Vector Network Analyzer (VNA).
During this simulation, the distances of the feedhorn and the probe to the pupil are held fixed, while the probe is moved 
perpendicular to the optical axis at regular steps. The forward and reverse transmission coefficients, $S_{21}$ and $S_{12}$, 
are calculated, and once they are normalized they can be plotted as a function of the scanning position to be compared with 
the measured normalized NF diffraction pattern. This comparison will be discussed later in Sect.~\ref{sec:scan}. 

During the numerical simulations we found that diffraction by the edges of the disk where the circular aperture is 
realized could cause anomalous amplitudes along the optical axis. Since this effect is not observable off-axis, we 
think that this may be a consequence of the circular symmetry of the system causing the diffracted waves from the disk
edge to interfere on the optical axis. This effect is geometry-dependent, and thus we chose the radius of the disk and the 
distance of the probe from the disk so that we did not detect this effect.

\section{EXPERIMENTAL RESULTS}
\label{sec:exp}

\subsection{Preliminary tests}
\label{sec:prelim}

Our experimental measurements had two main goals: 
{\it (a)} detect and quantify the  super-resolution effect with at least two TPs having 
different geometrical shapes; and, {\it (b)} evaluate and possibly reduce some of the effects that can
mask and/or alter the super-resolution effect.  An additional goal consisted in the determination 
of the FF patterns from NF measurements.  In fact, for a circular aperture 9\,cm in diameter, 
such as the one used in our tests, the Fraunhofer distance,
$2\,D^2/\lambda$, is about 1\,m. We therefore decided to perform NF measurements instead
of FF measurements for several reasons:  {\it (i)} the Fraunhofer distance is of the same
order of magnitude as the length of the anechoic chamber (about 7\,m); {\it (ii)} FF
scanning techniques would require more complex and expensive mechanical and microwave equipment,
as well as measuring methods; {\it (iii)} our preliminary tests indicated that we did not have
enough sensitivity to make measurements of the diffracted FF at distances of a few meters;
{\it (iv)} the FF would be severely affected by diffraction effects caused by the ground
screen supporting the TP, unless a much larger screen was adopted.

We note that these requirements are specific to our laboratory setup. On a telescope the TP,
like any other optical device designed to modify the incident plane wavefront, should ideally operate at the 
entrance pupil of the telescope, i.e., the primary mirror. For antennas and telescopes where this is not possible, 
an image of the entrance pupil can be used to place a transmittance filter~\cite{olmi2017}. 
%

\subsubsection{Measurement setup}
\label{sec:setup}

\begin{figure}  
   \begin{center}
   \begin{tabular}{c} 
   \hspace{-0.5cm}
   \includegraphics[height=5.0cm]{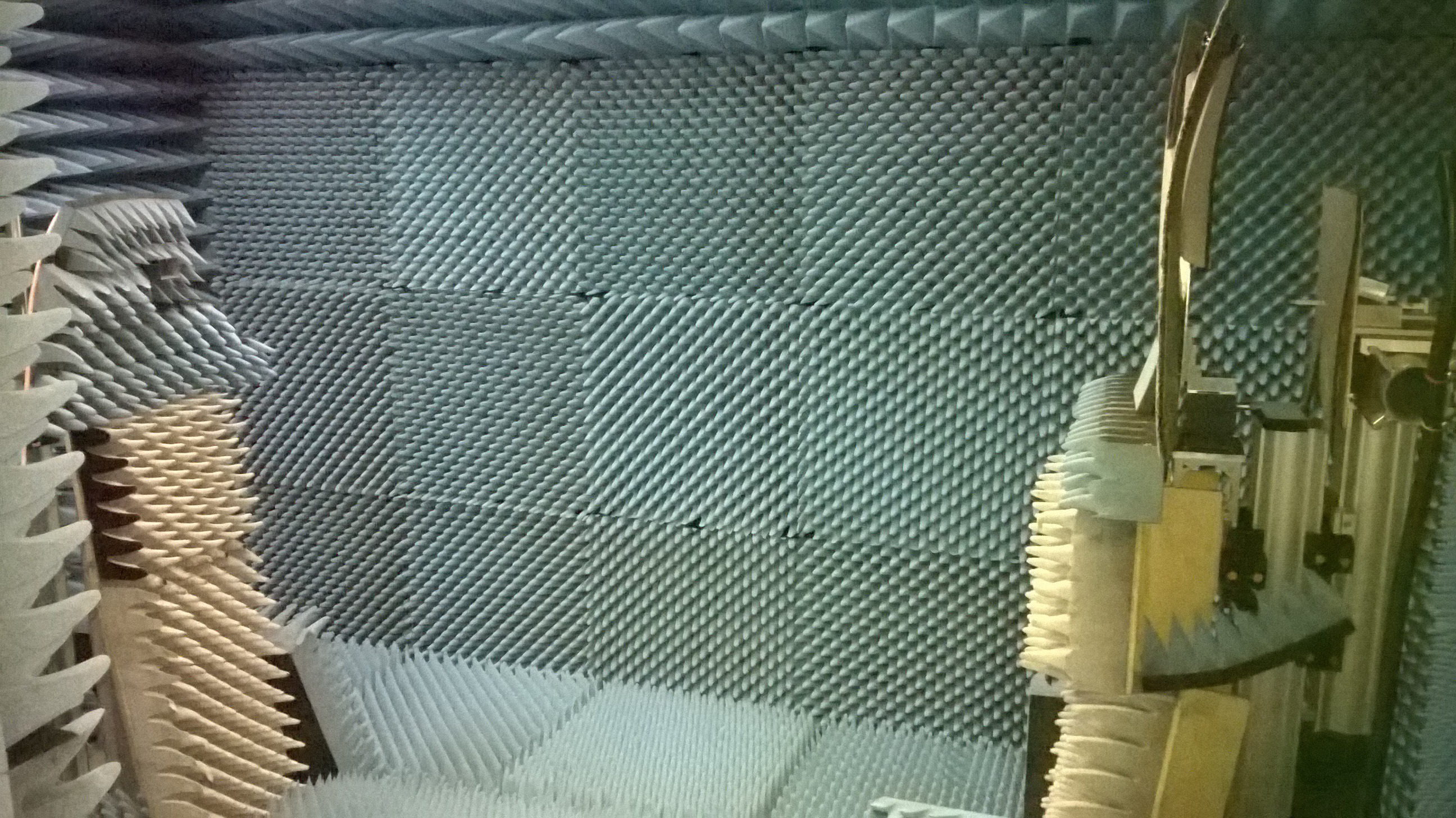}           
   \end{tabular}
   \end{center}
   \caption[example]
   { \label{fig:lab1}
Experimental setup for our laboratory tests at $\nu = 20\,$GHz. 
The launcher (rectangular feedhorn) is visible to the left, surrounded by absorbing cones.
On the right one can see the screen with the circular hole holding the TP 
and behind it the NF probe mounted on a linear stage 
to scan the field on an axis perpendicular to the direction of propagation, 
$z$. Later, the linear stage has been replaced with a two axis translation stage.
All components are mounted on vertical supports that can be
moved along $z$ on an optical bench set on the ground, which is covered with absorbing panels in this picture.
}
   \end{figure}

All microwave components were mounted on vertical supports that allowed to raise the optical 
axis about 1.5\,m above the ground, in order to minimize the effects of reflections
and multiple scatterings from the ground and the optical bench, though covered with 
microwave absorbers. The optical bench, 2.5\,m long (see Fig.~\ref{fig:lab1}) allowed all 
vertical supports to move along the direction of propagation, which we will indicate as the $z$-axis.

\begin{figure}  
   \begin{center}
   \begin{tabular}{c} 
   \includegraphics[height=10cm]{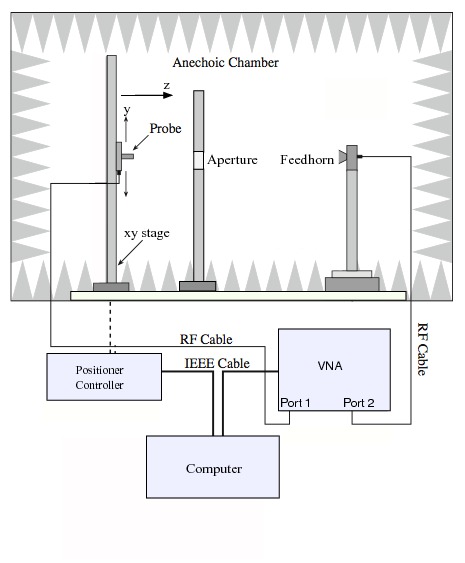}            
   \end{tabular}
   \end{center}
   \caption[example]
   { \label{fig:lab1schematic}
Schematics of the NF measurement system.  
}
   \end{figure}

All measurements were performed with a VNA  (Anritsu 37277C) recording both amplitude and phase.
Coaxial cables connected the VNA to the launcher, a rectangular feedhorn having a 
mouth $4 \times 5.5\,$cm in size, while the waveguide had dimensions $1.07 \times 0.43\,$cm 
with a total length of 11\,cm. The horn had a taper of $-3$\,dB at an angle 
of $\simeq 9\,$deg (FWHM $=18\,$deg) with respect to the optical axis, and a gain of $\simeq 20\,$dB.
The NF probe consisted of a section,  18\,cm in length, of an open-ended waveguide WR42 with smooth edges. 
Both the feedhorn and the probe were aligned so that their $E$-field was directed vertically, or along the $y$-axis,
according to Fig.~\ref{fig:lab1schematic}.
The probe was mounted on a manual translational stage to scan the NF
on an axis perpendicular to the direction of propagation (see Fig.~\ref{fig:lab1}). In between the feedhorn 
and the probe was mounted an assembly supporting a metal ground screen with a circular aperture 
which allowed to mount and remove the TPs. This assembly was covered by thin ($\simeq 1$\,cm thick)
planar microwave absorbers on the side facing the feedhorn. A schematics of the test set-up is shown 
in Fig.~\ref{fig:lab1schematic}.

After warm-up, both phase and amplitude measured with the VNA remained quite stable with 
peak-to-peak variations $\la 0.1\,$dB for the amplitude and $\la 0.5\,$deg for the phase.
The feedhorn and the probe were first aligned using a HeNe 
laser positioned behind the horn.  The circular aperture with the TP was then added to the 
optical bench and aligned as well. The coplanarity of the mouth of the horn and the plane of
the circular aperture was checked using a retroreflector.


\subsubsection{Noise measurements}
\label{sec:noise}

\begin{figure}  
   \begin{center}
   \begin{tabular}{c} 
   \hspace{-1.0cm}
   \includegraphics[height=5.5cm]{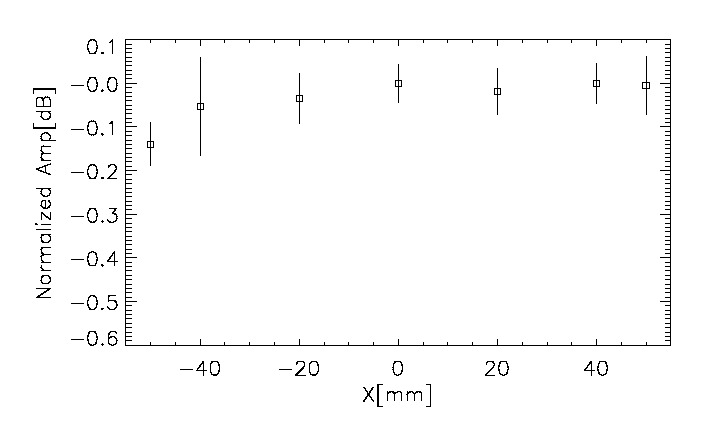} \\ 
   \hspace{-1.0cm}
   \includegraphics[height=5.5cm]{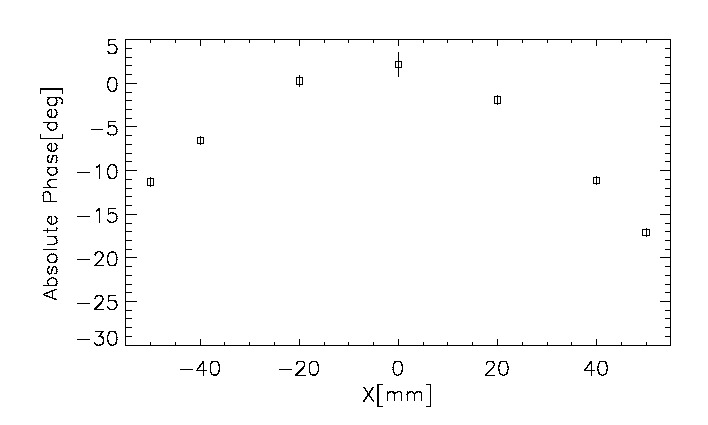}    
   \end{tabular}
   \end{center}
   \caption[example]
   { \label{fig:amp2m}
{\it Top panel.}
Measured normalized amplitude at $\nu = 20\,$GHz as a function of position along the horizontal direction, $x$,
perpendicular to the optical axis (located at $x=0$). The probe was set at a distance along $z$ of 2\,m from the feedhorn.
{\it Bottom panel.}
Same as the top panel for the phase. The error bars have roughly the same size as the symbol. 
}
   \end{figure}

After the initial setup, we proceeded to measure the RF alignment and the noise level, both in amplitude and phase. 
We carried out these tests with just the launcher and the probe. 
The probe was decentered manually
to scan the field from -5 to +5cm in the horizontal direction, $x$, perpendicular to the optical axis, $z$. The horizontal scan was
repeated five different times, and the mean and standard deviation of both amplitude and phase were estimated for each 
probe position (see Fig.~\ref{fig:amp2m}). The amplitude is quite uniform along $x$, with peak-to-peak variations
of less than 0.15\,dB. 

The power pattern of the feedhorn is expected to decrease as $P/P_o = \exp[-0.5(\theta/\sigma)^2]$, 
where $P_o$ represents the on-axis power level, $\theta$ is the angle of observation measured with respect to 
the optical axis and $\sigma = \theta_{\rm FWHM}/2.3548$.
Thus, the expected power (or amplitude) variation at $x=\pm 5\,$cm is $-0.002\,$dB, and we note that the observed amplitude variation 
along the $x$-axis in the top panel of Fig.~\ref{fig:amp2m} is well within the error bars of the single measurement. 
These non-repeatable errors are due to various sources of scattering within the anechoic chamber which limit our ability 
to measure amplitude variations $\la 0.05-0.1\,$dB. Further tests have shown that we can appreciate amplitude variations 
($\ga 0.1\,$dB) along the scan axis if the separation between the feedhorn and the probe is $\la 1.8\,$m.
However, the bottom panel of Fig.~\ref{fig:amp2m} shows that we can measure
phase variations with greater accuracy. At a distance of $\simeq 2\,$m the spherical wavefront propagating from the feedhorn 
should cause a phase variation of about 11\,deg at $x=\pm 5\,$cm, and this is approximately what we see in 
Fig.~\ref{fig:amp2m}. The slightly different phase variation at $x=-5\,$cm and  $x=+5\,$cm is a measure of the RF misalignment.

%
%

\subsection{Linear scanning in the near-field}
\label{sec:scan}

\begin{figure}  
   \begin{center}
   \begin{tabular}{c} 
\hspace{-1.0cm}
    \includegraphics[height=5cm]{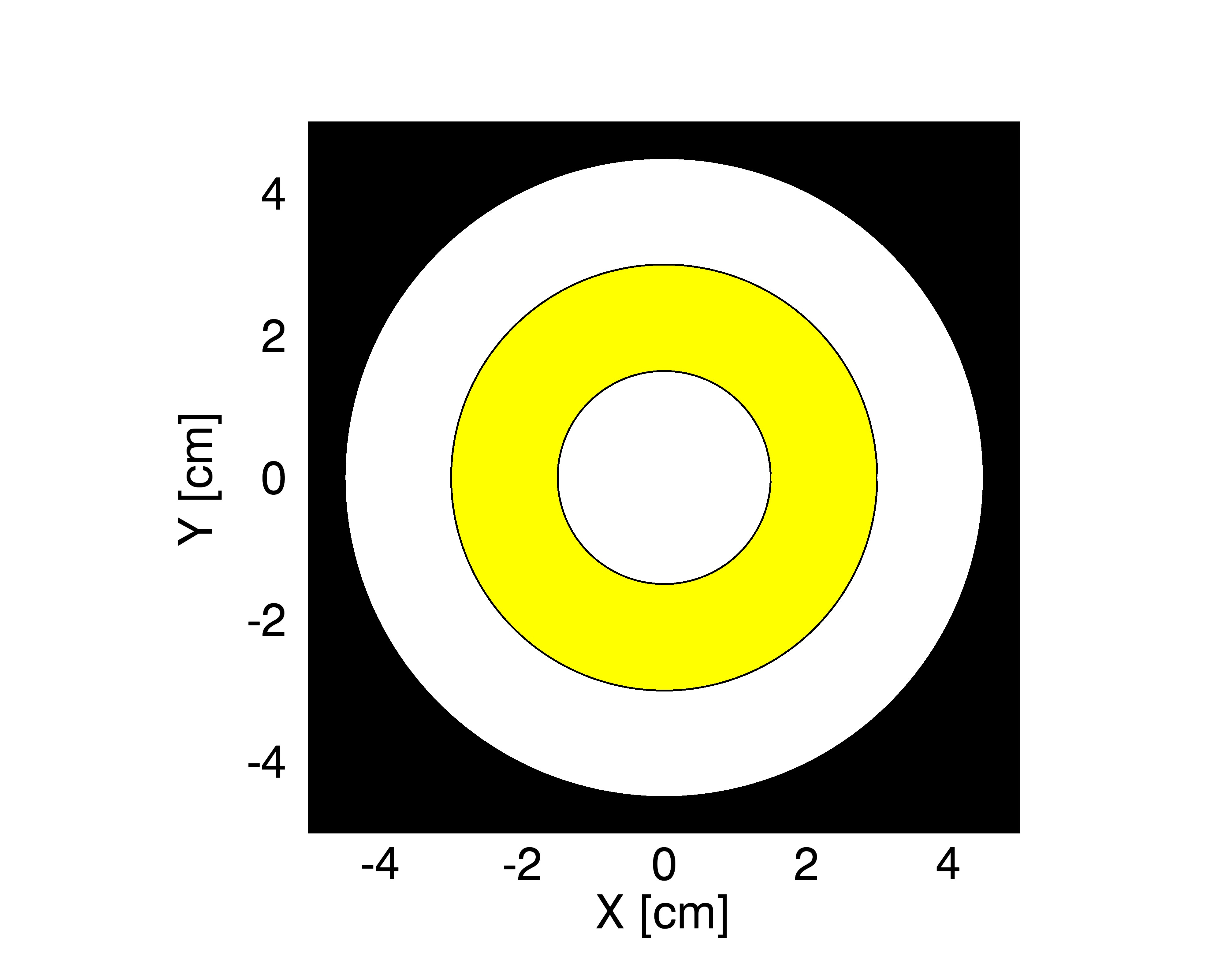}             
\hspace{-1.0cm}
    \includegraphics[height=5cm]{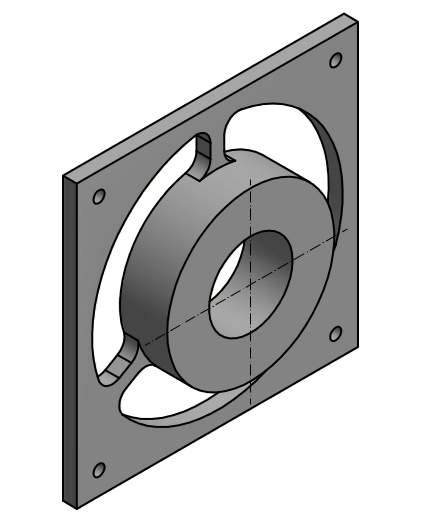}  \\       
\hspace{-1.3cm}															
    \includegraphics[height=5cm]{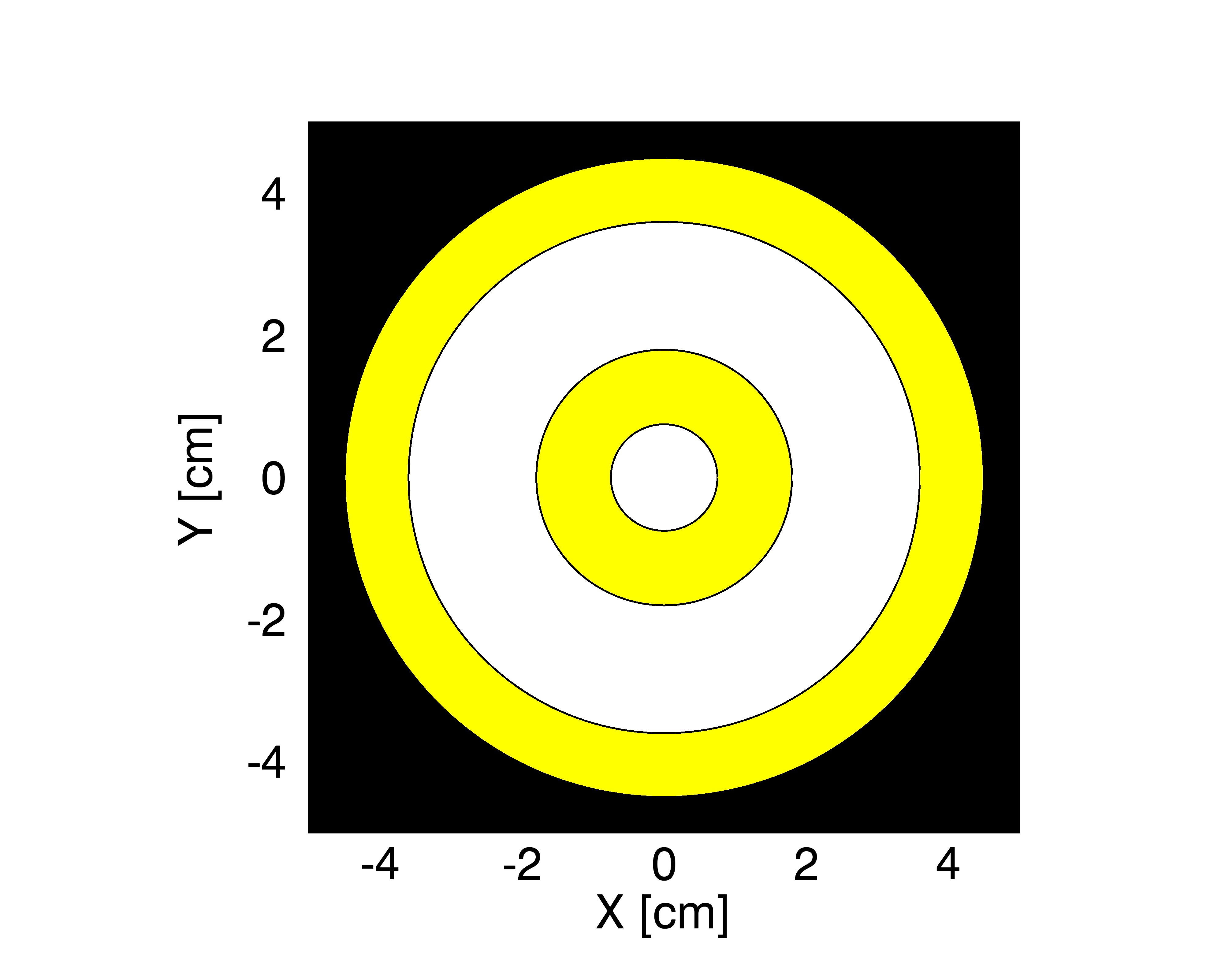}             
\hspace{-1.0cm}
    \includegraphics[height=5cm]{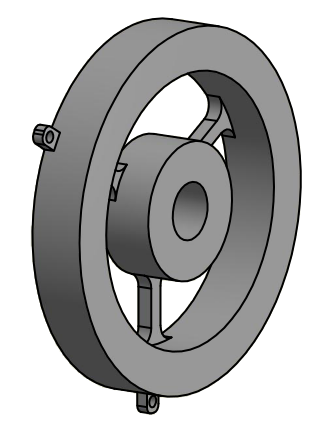} 
   \end{tabular}
   \end{center}
   \caption[example]
{ \label{fig:tp3}
{\it Top panels.}
Schematics of a TP with three-coronae ({\it left}), whose radii are 1.5, 3.0 and 4.5\,cm. The corona  in the middle (in yellow) 
is the one providing the phase inversion. In the right panel we show the CAD drawing used for the fabrication of this TP using
low-loss polyethylene. The thickness of the cylinder required to achieve the phase inversion is 1.5\,cm. 
{\it Bottom panels.}
Same as above for the TP with four-coronae. The radii of the four coronae are 0.75,  1.55, 3.60    and 4.50\,cm.  
}
   \end{figure}

After the preliminary tests were completed we proceeded with the measurements of the diffracted fields
by the open circular pupil and the same three- and four-coronae TPs described 
in Ref.~\cite{olmi2016}. Figure~\ref{fig:tp3} shows the layout of these TPs, which were 
fabricated from low-loss polyethylene, having a relative dielectric constant $\epsilon_r = 2.28$ and a loss tangent
of $\tan \delta \simeq 3.8\,10^{-4}$~\cite{goldsmith1998}.

\begin{figure}  
   \begin{center}
   \begin{tabular}{c} 
   \hspace{-1.0cm}
   \includegraphics[height=5.5cm]{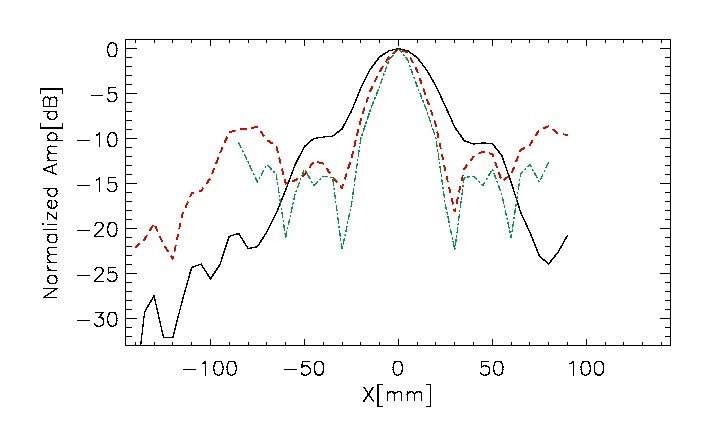}  \\   
   \hspace{-1.0cm}
   \includegraphics[height=5.5cm]{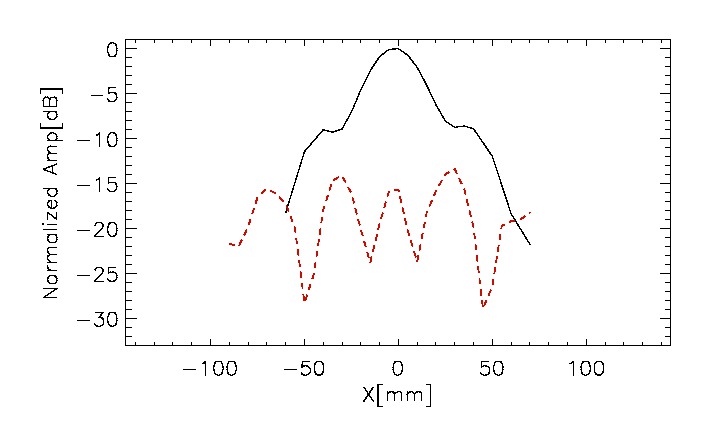} 
   \end{tabular}
   \end{center}
   \caption[example]
{ \label{fig:linscanTP3}
{\it Top panel.}
Plot of the amplitude at 20\,GHz in the NF of the open pupil (solid, black line), and of the three-coronae TP
shown in Fig.~\ref{fig:tp3} (dashed, red line). More measurements were performed on just one side of the scan 
(the negative $x$ direction), in order to test the strength of the far sidelobes. 
For comparison, the simulated amplitude distribution from FEKO,
as discussed in Sect.~\ref{sec:spar}, is also shown (dot-dashed, green line). 
All curves are normalized with respect to the peak value. 
The distance between the launcher and the screen was 1.9\,m, the separation between the screen and the probe was 18\,cm, 
and the sampling interval along the scan direction was 0.5\,cm.
{\it Bottom panel.}
Same as above for the four-coronae TP.  The distance between the launcher and the screen was 1.7\,m, the separation 
between the screen and the probe was 16\,cm, and the sampling interval along the scan direction was 0.5\,cm. In  this case 
the amplitude generated by the TP is normalized to the peak value of the open pupil. 
}
   \end{figure}

A scan along the $x$-direction was first performed for the open pupil, roughly between $x=-10\,$cm and $x=+10\,$cm, with 
the probe positioned at different distances from the screen. In the top panel of Fig.~\ref{fig:linscanTP3} we show an example 
of our results obtained using the three-coronae TP 
with a distance between the launcher and the screen of 1.9\,m, while the separation between the screen and the probe was 
18\,cm. Even a simple linear scan like this clearly shows that the FWHM of the amplitude distribution generated by
the TP is smaller than that of an open pupil with the same diameter, and thus we detect the super-resolution effect,
at least in the NF.  Later we will show that this relation also holds in the FF. We note that the open pupil and
TP fields have been measured  more extensively on the negative $x$-direction. This extended scan was intended
to record the sidelobes level and also to check the measurement setup with low signal levels. Several combinations
of distance between the launcher and the screen, and separation between the screen and the probe were experimented. The
bottom panel of Fig.~\ref{fig:linscanTP3} was in fact obtained using a different set of parameters.

%

The top-panel of Fig.~\ref{fig:linscanTP3} also shows the radiation diagram obtained with FEKO using the
$S$-parameters method described in Sect.~\ref{sec:spar} (the few points shown are due to the long time
required by each separate simulation). We note that the simulated pattern
has a close correspondence with the measured  signal, particularly regarding the positions of the nulls and 
the first sidelobes. Both diffraction patterns are sparsely sampled and thus part of the discrepancies 
between the two curves is likely a result of the insufficient sampling. 
It should also be noted that the radiation diagram produced by the TP in Fig.~\ref{fig:linscanTP3}
also includes the effects of the three small rods that are used to hold the dielectric cylinder of the TP connected to the 
external rim, as shown in the right panel of Fig.~\ref{fig:tp3}. However, we have performed FEKO simulations including 
the rods and we have verified that the diffraction effects introduced by these supports are negligible.

%
   \begin{figure}  
   \begin{center}
   \begin{tabular}{c} 
   \hspace{-1.0cm}
   \includegraphics[height=5.5cm]{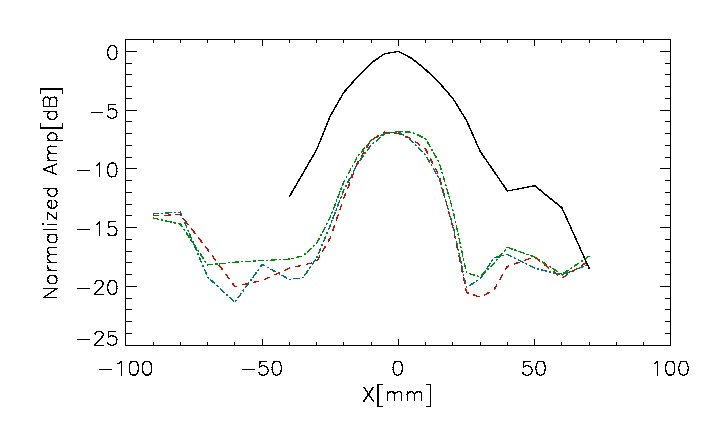}     
   \end{tabular}
   \end{center}
   \caption[example]
{ \label{fig:linscanTP3freq}
Frequency response of the three-coronae TP. The NF of the open pupil is also
represented by the solid, black line, while the
dashed and dot-dashed lines represent the fields of the TP at 19.84, 20.0 and 20.24 GHz,
normalized with respect to the peak value of the open pupil so the actual amplitude decrease
is visible. The launcher and probe distance from the pupil plane were
1.7\,m and 20\,cm, respectively, while the sampling interval was 1.0\,cm instead of
0.5\,cm as in Fig.~\ref{fig:linscanTP3}.
}
   \end{figure}

The bottom panel of Fig.~\ref{fig:linscanTP3} shows the radiation diagram obtained with the four-coronae TP. 
Since in this case the amplitude generated by the TP is normalized to the peak value of the open pupil, one can
appreciate the general decrease of the signal level in the NF when the TP is mounted on the circular aperture.
While the central lobe shows the super-resolution effect, the relative level of the sidelobes is much higher compared
to the three-coronae TP. Note that the individual FWHM cannot be compared between the top and the bottom panels
because the separations between launcher, pupils and probe were different.

Finally, a critical question to be analyzed is the frequency response of a TP, which is important in the 
perspective of its possible use on a radio telescope. 
A wide-bandwidth analysis of the TPs was out of the scopes of this work. However, we
have performed a simple test by measuring the
response of the three-coronae TP at three separate frequencies, specifically 19.84, 20.0 and 20.24 GHz, thus spanning
a 400\,MHz range. The results are shown in Fig.~\ref{fig:linscanTP3freq}, where all curves are
normalized with respect to the peak value of the open pupil so the actual amplitude decrease when using the TP is visible. 
The launcher and probe distance from the pupil plane were slightly different from the example discussed above, but this 
has no effect on the present discussion. We note that the main lobes at the three frequencies are quite similar, while
most of the differences are limited to the sidelobes. The sampling interval along the scan direction was larger
compared to the previous example, and this may also contribute to some of the discrepancies, especially at low signal levels.
These measurements suggest that the three-coronae TP has a usable bandwidth $\ga 400\,$MHz. Given that most
K-band receivers on currently operating radio telescopes have instantaneous bandwidths $\sim 1-2\,$GHz, a TP 
device might be partially limiting the available bandwidth.

%
%
   \begin{figure*}  
   \begin{center}
   \begin{tabular}{c} 
   \hspace{-1cm} 
   \includegraphics[height=10cm,angle=0]{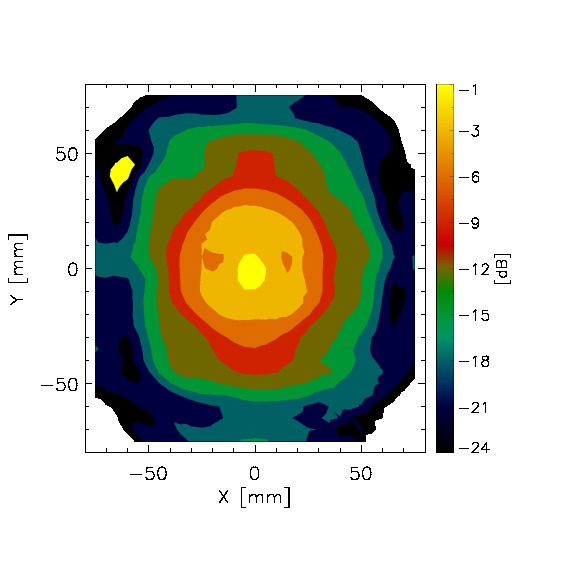}        
    \hspace{-1cm}                                          
   \includegraphics[height=10cm,angle=0]{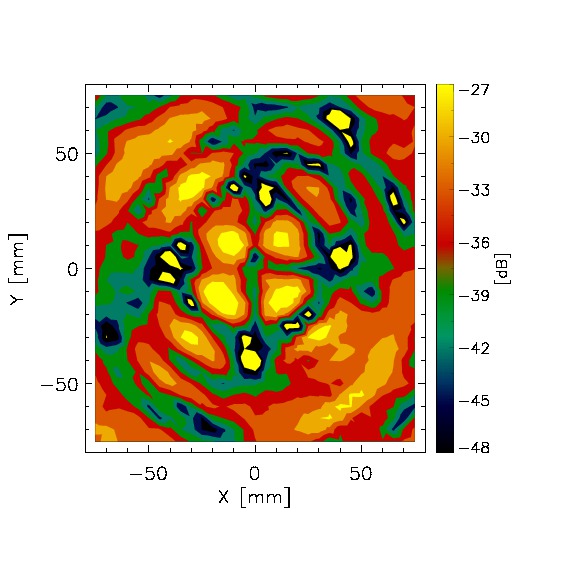}               
   \end{tabular}
   \end{center}
   \caption[example]
{ \label{fig:open2D}
{\it Left panel.} 
Near-field copolar amplitude distribution at 20\,GHz  generated by an open circular 
pupil with diameter 9\,cm, at a distance of 10\,cm from the plane of the pupil. 
The launcher was at a distance of 1.7\,m from the plane of the pupil. 
The high-signal spot in the top-left of the map is likely to be an artifact, 
since it does not show up in subsequent maps.
{\it Right panel.}
Same as the left panel for the cross-polar component. The maps are normalized 
to the peak amplitude value of the copolar pattern.  
}
   \end{figure*}

\subsection{Planar scanning setup}
\label{sec:planarsetup}

After the preliminary measurements were completed, 
the NF probe was mounted on a motorized, two axes translation stage, which could be 
remotely controlled to move the probe at the required position in the $x-y$ plane,  
thus performing a {\it planar scanning} of the NF with 
uniform sampling step (plane-rectangular scanning).  Likewise, the feedhorn was mounted on a 
rotational stage, a feature that was required to precisely rotate the horn
and measure the cross-polarization.  

In preparation for the measurements, the RF absorbing panels were appropriately positioned all around 
the metallic structure of the 2D positioner. 
The usual optical and RF alignment were then performed. 
Both VNA and translation stage were remotely controlled through
{\tt LabVIEW}\footnote{\tt http://www.ni.com/labview/}. The operator would specify the $x$-$y$ size and step 
of the planar scanning to be performed by the probe in the NF and the control software would then 
move the probe at each position in the grid, where it would stop while the 
VNA measurement is acquired, and then would move to
the next position.  A typical raster map with $31 \times 31$ points and a 0.5\,cm sampling interval would take 
$\simeq 1\,$hr to complete. The complex voltage samples are stored together with their positions and will 
later be used for the NF-to-FF conversion (see Sect.~\ref{sec:nf2ff}). 

\subsection{Measurements results}
\label{sec:meas}

%
%
   \begin{figure*}  
   \begin{center}
   \begin{tabular}{c} 
   \hspace{-1cm}
   \includegraphics[height=10cm,angle=0]{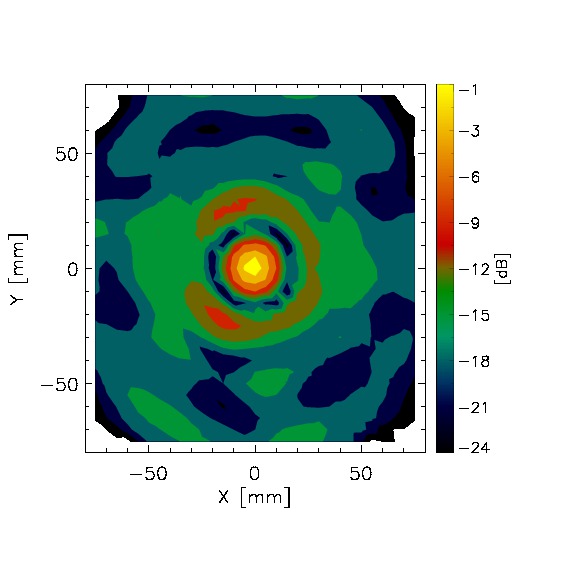}
    \hspace{-1cm}
   \includegraphics[height=10cm,angle=0]{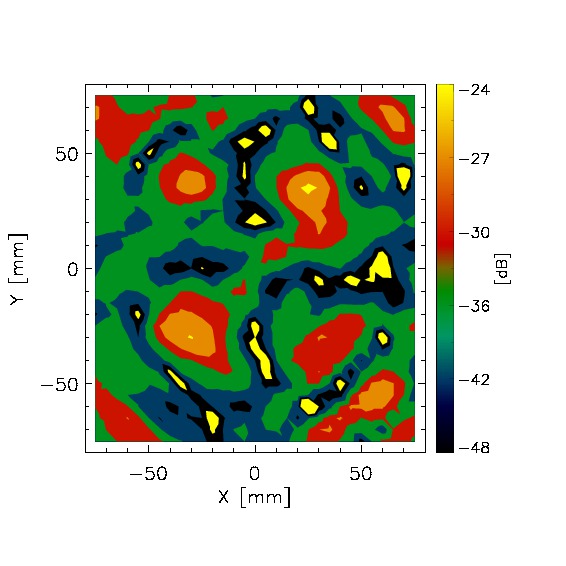}
   \end{tabular}
   \end{center}
   \caption[example]
{ \label{fig:TP32D}
Same as Fig.~\ref{fig:open2D} for the three-coronae TP.
}
   \end{figure*}

%
%
   \begin{figure*}  
   \begin{center}
   \begin{tabular}{c} 
   \hspace{-1cm}
   \includegraphics[height=10cm,angle=0]{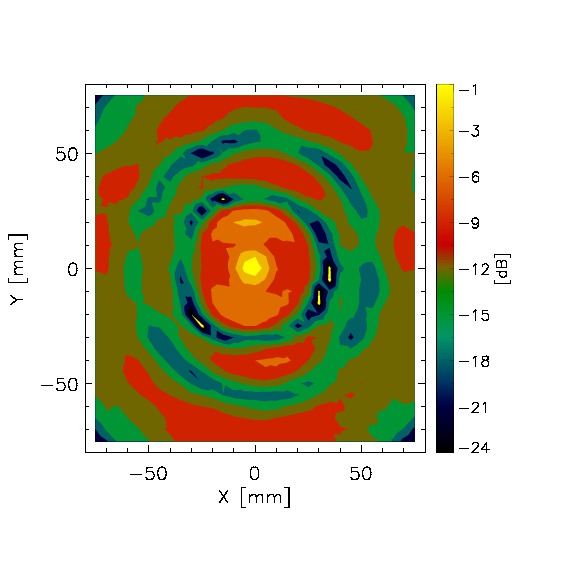}
    \hspace{-1cm}
   \includegraphics[height=10cm,angle=0]{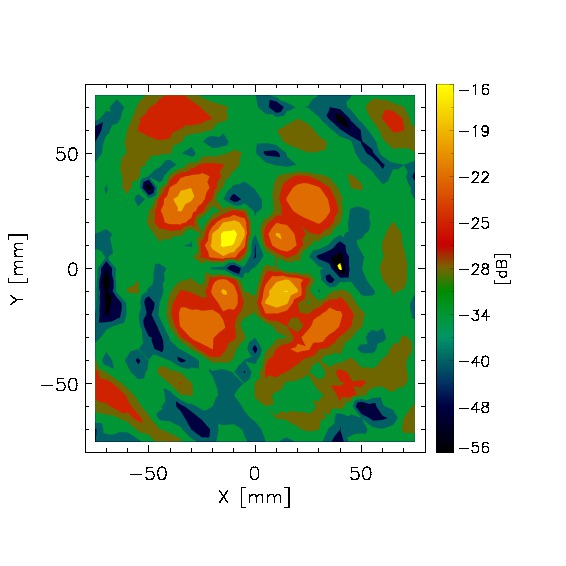}
   \end{tabular}
   \end{center}
   \caption[example]
{ \label{fig:TP42D}
Same as Fig.~\ref{fig:open2D} for the four-coronae TP.
}
   \end{figure*}

Raster maps were performed with the open pupil and with both three- and four-coronae TPs. 
If the incident field over the aperture is linearly polarized (see Section~\ref{sec:setup}), 
then in the scalar diffraction approximation (see Section~\ref{sec:analytical}) 
we need to consider only the scattered fields 
along the polarization direction~\cite{silver}. However, because we also required to perform 
the NF-to-FF transformation (see Sect.~\ref{sec:nf2ff}), both our measurements and EM
simulations had to determine the copolar and cross-polar\footnote{Although the copolar and  cross-polar
terms are generally referred in the literature to the FF, here we will use them to indicate the vertical, $y$-component,
and horizontal,  $x$-component,  of the NF.}
radiation patterns of the scattered fields.
In addition, the comparison of copolar and cross-polar components also had an interest in 
preparation to a potential application of TPs with the receiver system of a radio telescope.

The cross-polar pattern was recorded after rotating the feedhorn by 90\deg.
The major challenge for the measurement of the cross-polar pattern was clearly
the low-level of the signal, since we were approaching the expected dynamic range limit of our experimental setup. For this 
reason we selected the distance between the feedhorn and the pupil, and the distance between the pupil and the probe to achieve
an acceptable trade-off in terms of planarity of the incident wave and signal levels, that would allow the cross-polarization
to be measured. We performed planar scannings using various distances between the probe and the pupil. However, in the
end we selected a separation of 10\,cm because larger separations would also lead to a larger truncation error (see Sect.~\ref{sec:nf2ff}).

The copolar and cross-polar amplitude patterns of the open circular pupil are shown in Fig.~\ref{fig:open2D}. The copolar
pattern has a slightly elliptical shape which, however is not visible in Fig.~\ref{fig:TP32D} and is only partially visible 
in Fig.~\ref{fig:TP42D}. We do not have enough data to further investigate  this issue, which might be a consequence 
of the overall accuracy of the measurement procedure.
The cross-polar pattern also shows 
a good signal-to-noise ratio (SNR), which allows to measure its level at least 27\,dB below the copolar signal. 
Achieving a good SNR is a necessary condition in order to allow a reliable  
reconstruction of the FF. As a repeatability test, we have performed this map again after several 
days and found that the shape of the pattern was essentially the same, confirming the robustness of the measurements setup. 

Next, we have performed the raster map with the three-coronae TP mounted on the circular aperture. 
The copolar and cross-polar amplitude patterns are shown in Fig.~\ref{fig:TP32D}. The left panel shows a nicely circular 
main lobe and a first sidelobe with also an approximately circular symmetry, which  indicate a good overall RF alignment. 
The cross-polar map in the right panel of Fig.~\ref{fig:TP32D} also shows the typical four-lobe pattern, but with clearly lower 
SNR compared to Fig.~\ref{fig:open2D}. It should be noted, in fact, that the 0\,dB signal level is referred to the peak value 
of the copolar component of the TP and not of the circular aperture, and thus the absolute amplitudes in Fig.~\ref{fig:TP32D} are 
many dBs below the levels of Fig.~\ref{fig:open2D}. In addition, the copolar pattern has a FWHM that is clearly smaller than that
of the open pupil, which confirms the preliminary results described in Sect.~\ref{sec:scan}.
 
Finally, the NF maps obtained with the four-coronae TP are shown in 
Fig.~\ref{fig:TP42D}. Compared to Fig.~\ref{fig:open2D} the super-resolution effect is clearly visible, but a 
higher degree of asymmetry can be seen in the copolar pattern, which also shows higher sidelobes compared to 
the three-coronae TP. Higher sidelobes were already measured during the preliminary linear scans, as shown in
Fig.~\ref{fig:linscanTP3}. The cross-polar pattern is also clearly detected with the same symmetry as in Figures~\ref{fig:open2D} 
and \ref{fig:TP32D}.

\begin{figure}  
   \begin{center}
   \begin{tabular}{c} 
   \hspace{-1cm} 
   \includegraphics[height=7cm,angle=0]{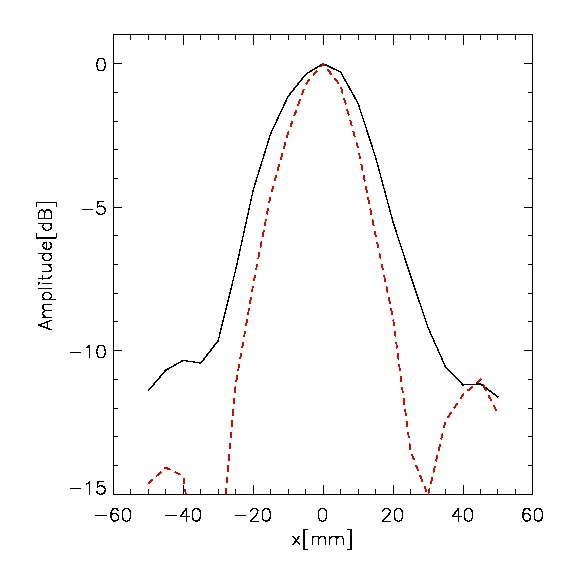}    \\
    \hspace{-1cm}                                          
   \includegraphics[height=7cm,angle=0]{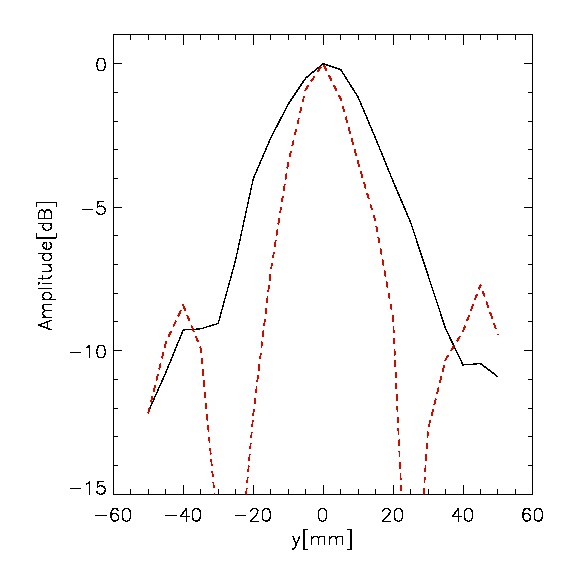}                                     
   \end{tabular}
   \end{center}
   \caption[example]
{ \label{fig:cutxy}
1D cuts of Figures~\ref{fig:open2D} and \ref{fig:TP32D} along the $x$- ({\it top panel}) and $y$-axis ({\it bottom panel}). 
Both amplitudes are normalized to their peak values.
}
   \end{figure}

In Fig.~\ref{fig:cutxy} we also show the transversal cuts, along the $x$- and $y$-axis, of the 2D amplitudes shown
in Figures~\ref{fig:open2D} and  \ref{fig:TP32D}. These curves should be compared with the top panel of
Fig.~\ref{fig:linscanTP3}. Despite the difference in the FWHM of the $x$- and $y$-cuts of the amplitude 
of the open pupil, due to the slight asymmetry discussed earlier, these results show otherwise a good 
agreement with the linear scan shown in Fig.~\ref{fig:linscanTP3}.

%
%

%
   \begin{figure*}[ht]
   \begin{center}
   \begin{tabular}{c} 
   \hspace{-1cm}
   \includegraphics[height=6cm,angle=0]{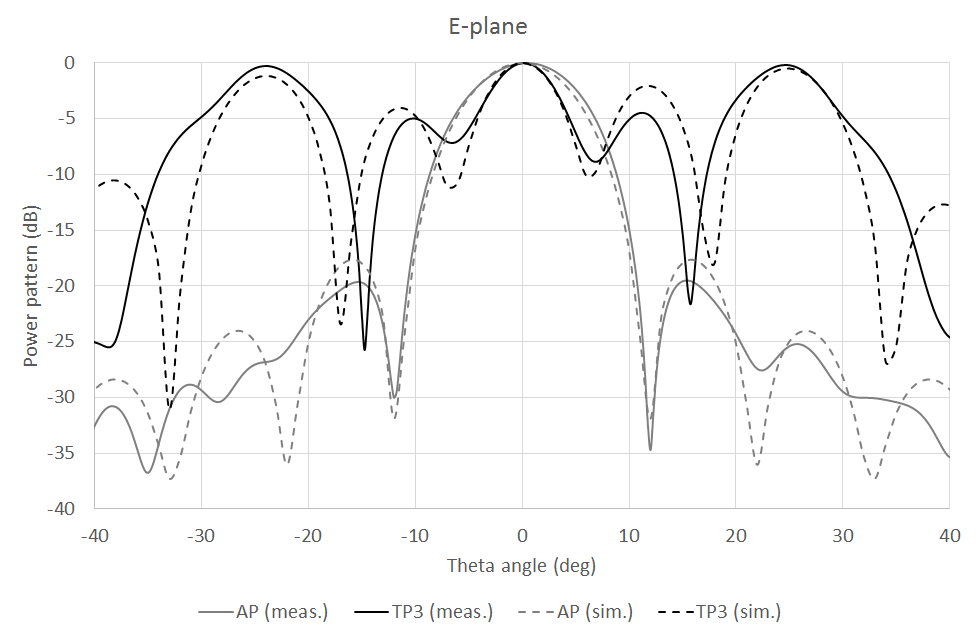}   
   \includegraphics[height=6cm,angle=0]{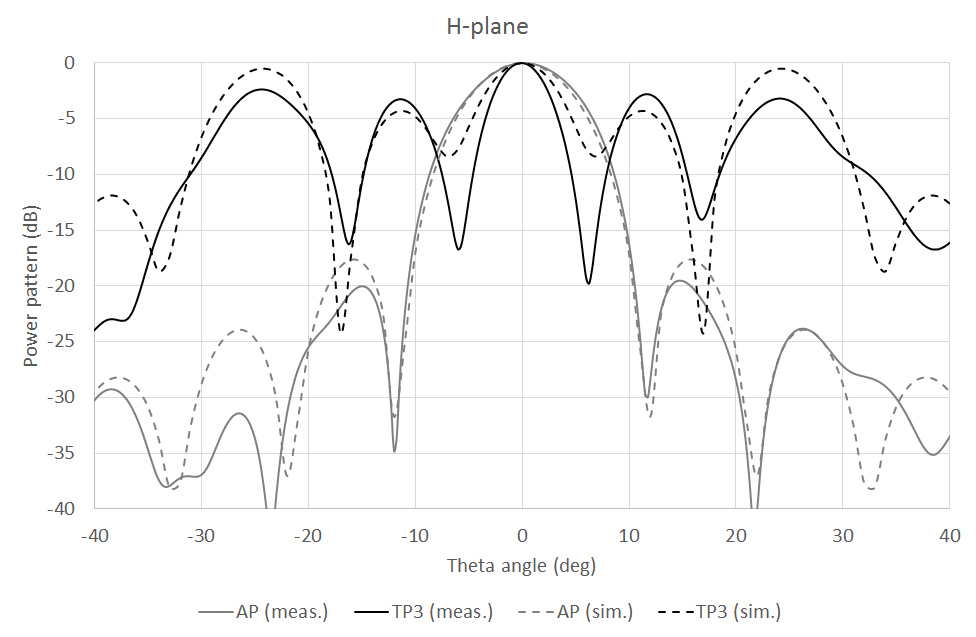}
   \end{tabular} 
   \end{center}
   \caption[example]
{ \label{fig:PT3FF}
Normalized power patterns (from $-40$ to 0\,dB) obtained in the case of the three-coronae TP 
in the $E$-plane ({\it left panel}) and $H$-plane ({\it right panel}), for an angular range $-40^{\circ}$ to $40^{\circ}$
(the angle $\theta$ refers to the angular separation between the optical axis and the direction of observation).
The solid lines represent the measured data (after the NF-FF transformation),
while the dashed lines represent the FEKO models.
The radiation diagrams of the open pupil 
(indicated as APerture in the figures)
are shown for reference  (light-grey curves). 
}
   \end{figure*}

%
   \begin{figure*}[ht]
   \begin{center}
   \begin{tabular}{c} 
   \hspace{-1cm} 
   \includegraphics[height=6cm,angle=0]{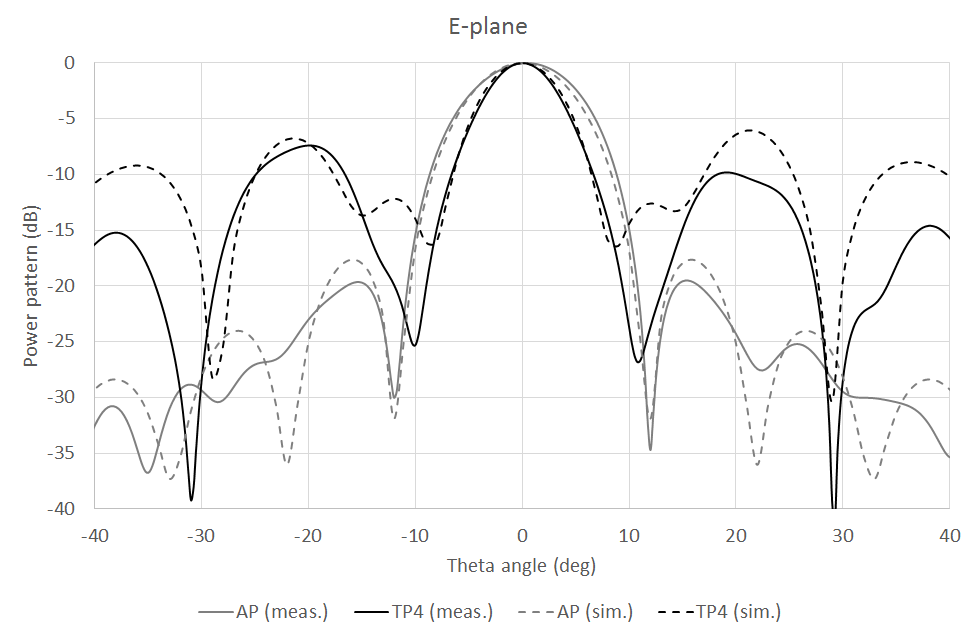}
   \includegraphics[height=6cm,angle=0]{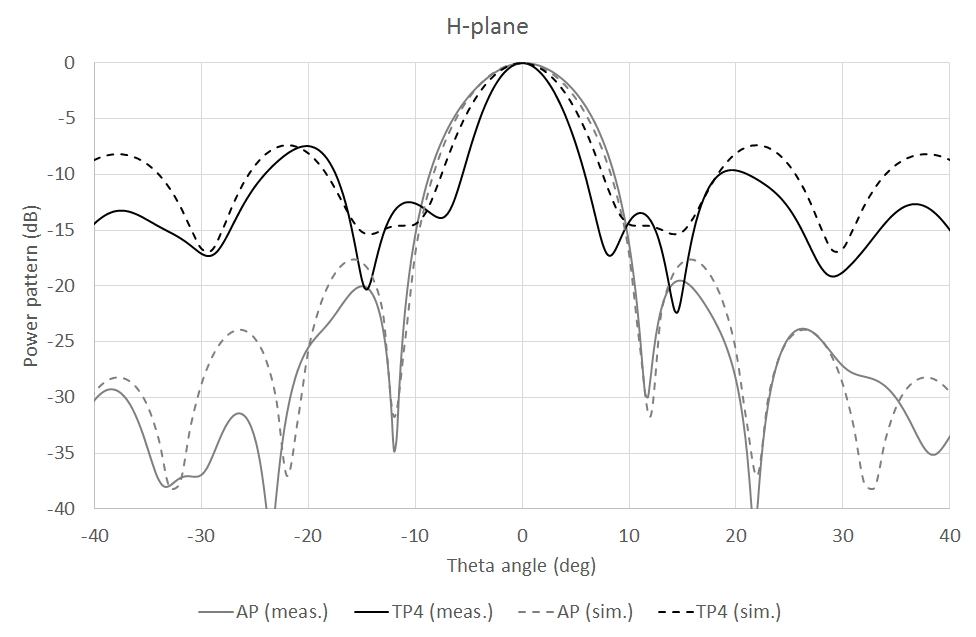}                                    
   \end{tabular}
   \end{center}
   \caption[example]
{ \label{fig:PT4FF}
Same as Fig.~\ref{fig:PT3FF} for the four-coronae TP.
}
   \end{figure*}

\section{NEAR-TO-FAR-FIELD TRANSFORMATION}
\label{sec:nf2ff}

Since all our measurements were performed in the NF of the open or composite 
circular pupil, a fundamental step in our work was 
the determination of the FF pattern from the NF  measurements. 
As it was earlier mentioned in Sect.~\ref{sec:planarsetup}, in preparation for our 
NF-FF transformation procedure, we employed a planar scanning technique with regularly 
spaced data (plane-rectangular scanning). This is certainly the simplest 
technique from the experimental and computational point of view, and in
Appendix~\ref{sec:method} we review the specific method used for the NF-FF transformation.
We apply the NF-FF transformation to the measured NF of the 
open and composite circular pupils, shown in Figures~\ref{fig:open2D} to \ref{fig:TP42D}. 

We show the FF of the three- and four-coronae TP in Figures~\ref{fig:PT3FF} and \ref{fig:PT4FF},
respectively, where they are also compared with the FF of the open pupil. We separately show the 
two components along the $E$ and $H$ reference planes, which correspond to cuts of the FF 
along the $y$- and $x$-axis, respectively (thus corresponding to the vertical and horizontal orientations
discussed in Appendix~\ref{sec:method}). The radiation diagrams show several interesting features. 
First of all, the super-resolution effect is clearly visible for both TPs, though at different levels, 
as expected. The FWHM is different along the two planes and is a consequence of 
the slight asymmetry of the radiation diagrams in the $H$- and $E$-planes.
Secondly, the number, intensity and position of the sidelobes is also different between the two TPs, 
which is also expected, as discussed in Sect.~\ref{sec:toraldo} (see also Ref.~\cite{olmi2016}).
In particular, the four-coronae TP has much lower level sidelobes, though the resolution gain, $G$, is
somewhat lower compared to the three-coronae TP. 
Finally, the normalized power patterns in Figures~\ref{fig:PT3FF} and \ref{fig:PT4FF} do not show 
that the radiation diagrams of the TPs lie several dBs below the intensity level of the open pupil.
Therefore, before TPs can be used effectively on a radio telescope their overall efficiency must
be increased (see also Section~\ref{sec:concl}).

When comparing the measured (and NF to FF transformed) FF of the TPs with the simulated FF from FEKO,
we note from Figures~\ref{fig:PT3FF} and \ref{fig:PT4FF} that the simulated FF of the open pupil 
closely follows the measured field, particularly in the $H$-plane. However, in the case of the TPs 
we can note some discrepancies between the measured and simulated fields, which are
more evident in the $H$-plane. At present, we do not have a convincing explanation of as to why 
in the $H$-plane we simultaneously observe a better agreement between the measured and simulated FF 
of the open pupil, and a larger discrepancy in the case of the composite pupils. A tentative explanation
could be that the dielectric coronae are not modeled with sufficient accuracy by our numerical simulations 
in FEKO. In all cases we can also note an increasingly larger discrepancy between the measured
and modeled sidelobes at larger observing angles.  This effect, however, is better known and is associated
with the truncation effect discussed in Appendix~\ref{sec:method}.

Given all uncertainties and limitations in both the experimental measurements and numerical simulations,
we think that the agreement between the measured and simulated FF is reasonably good, and confirms the
ability of a discrete TP to achieve the super-resolution effect even in the absence of the required amplitude
apodization, as previously discussed in Sect.~\ref{sec:toraldo}. These measurements also show that it
is indeed possible to achieve a trade-off between the super-resolution effect and the sidelobes level when 
a larger number of coronae is used in the design of the TP.

\section{CONCLUSIONS}
\label{sec:concl}

``Toraldo Pupils", or variable transmittance filters, introduced by G. Toraldo di Francia in 1952~\cite{toraldo1952a,toraldo1952b}, 
represent a viable technique to achieve super-resolution, i.e., an angular 
resolution beyond the classical diffraction limit,  in the microwave range. 
One of the most important characteristic of the TP is its simplicity and ease of fabrication.
In order to investigate the possibility to apply these techniques to filled-aperture radio telescopes, we have first
performed a series of extensive EM numerical simulations~\cite{olmi2016} at a frequency of 20\,GHz, which
represents a trade-off between the computing time required by the EM simulations and
the size and separations of the microwave components in units of wavelengths. 
The simulated FFs confirm that a partial super-resolution effect can be achieved even without 
 the amplitude apodization required by the ideal optical model. We have also used 
 additional EM simulations to more accurately model the launcher (rectangular feedhorn)
 and the spatial sampling in the NF by the probe.

We have then carried out laboratory measurements of the diffracted NFs by different TPs 
and compared them with the corresponding diffraction pattern of a circular open pupil.
We first performed a series of preliminary tests, mainly devoted to measure various sources 
of noise and scattering, and then carried out planar scannings of the NF,
measuring both the copolar and cross-polar components of the fields. 
The NF was then transformed into the FF and we detect the super-resolving effect in both ranges. Comparing
our results in the FF with the FEKO numerical simulations, we find in general a good agreement. 
Our sensitivity is good enough to measure the sidelobes, allowing us to compare the level 
and number of sidelobes for the two TPs under test.
Our measurements confirm the results of the first experiments in the microwave range~\cite{mugnai2003,ranfagni2004}
and, in particular, they show that the super-resolution effect is achieved with both three- and four-coronae discrete TPs. 
The different resolution gain, $G$, and sidelobes obtained with the two TPs confirm that
the number and geometry of the coronae can be used to achieve a trade-off between $G$ and the sidelobes
relative intensity and position.  This is important if a specific device including a TP should be designed to
operate on  a radio  telescope.

Overall, our investigation confirms the super-resolving TP proof of concept. However, before TPs can be used 
efficiently on a radio telescope several problems must be addressed and solved. 
The main problem is that a variable transmittance filter should ideally be placed at the entrance pupil of 
a filled-aperture telescope, i.e., the primary reflector for classical two-mirrors telescopes. As this is clearly 
impractical (unless an active primary surface is available), a viable option is to place the TP at an {\it image} 
of the entrance pupil. This image can be generated through the use of a collimator, which also couples the 
TP device to, e.g.,  an existing receiver of a radio telescope. This and other design issues will be discussed in a 
forthcoming paper that will describe the preliminary design of a prototype TP optical system 
to be mounted on the 32-m Medicina antenna in Italy\footnote{\tt http://www.med.ira.inaf.it/}~\cite{olmi2017}. 

Another important problem that needs further analysis is that the gain in angular resolution,
obtained by the reduction in size of the main lobe, may be offset by the increase in intensity of the sidelobes and the
decrease in the aperture efficiency (or antenna gain~\cite{olmi2007}) of the radio telescope.
By using pupils with an array of concentric annuli, or even a {\it continuous} TP (Olmi {\it et al.}, in prep.), 
the PSF can be tailored in such a fashion that one can get a narrow central beam surrounded by neighboring sidelobes
of low intensity. The sidelobe level and position with respect to the optical axis are important especially in mapping
applications. However, in this case an alternative solution can be the application of deconvolution algorithms to
remove the high sidelobes from the final map~\cite{roy2011}.
We think that the loss in the aperture efficiency measured in our laboratory experiments can be mitigated by the use of 
efficient global optimization algorithms in the design of super-resolving pupil filters, as well as by
the use of metamaterials.

\begin{acknowledgements}
We gratefully acknowledge the contribution of the Ente Cassa di Risparmio di Firenze (Italy)
for supporting this research.
We also wish to thank G. Cauzzi (INAF-OAA) for providing us with some of the essential mechanical 
components used in our laboratory measurements, L. Carbonaro (INAF-OAA) for helpful suggestions 
regarding the mechanical setup and A. Ignesti (CNR-IFAC) for valuable suggestions on how to
perform the measurements in the anechoic chamber.
\end{acknowledgements}

\appendix
\section{Plane-rectangular NF-FF transformation}
\label{sec:method}

\begin{figure}  
   \begin{center}
   \begin{tabular}{c} 
    \includegraphics[height=6cm,angle=0]{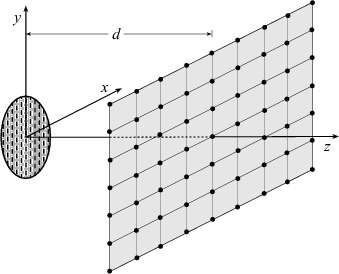}    
   \end{tabular} 
   \end{center}
   \caption[example]
{ \label{fig:2dscanning}
Schematics of the plane-rectangular scanning. 
}
   \end{figure}

Among the NF-FF transformation techniques (planar, cylindrical or spherical), 
the plane-rectangular scanning is the simplest and most efficient from the analytical 
and computational viewpoints. Such a technique is particularly 
suitable for highly directive antennas 
since the pattern can be reconstructed only in a cone with an apex angle less than $180^\circ$.
In the plane-rectangular scanning (Fig.~\ref{fig:2dscanning}), the probe is mounted on a $x$-$y$ 
positioner so that it can measure the NF amplitude and phase on a plane-rectangular grid. 
From these data, measured for two orthogonal orientations (horizontal, $H$, and vertical, $V$) 
of the probe, or equivalently the transmitting feedhorn 
(by applying a $90^\circ$ rotation around the longitudinal axis in the second set), and taking 
into account the probe spatial response, one can compute the antenna FF pattern~\cite{paris1978,joy1972}. 

%

It can be easily recognized that the NF tangential components of the field ($E_x$ and $E_y$) 
cannot be obtained when performing the measurement by means of a real probe. 
In fact, the probe sees the center of the diffracting pupil (which constitutes our
``antenna under test'', or AUT)
from different directions when moving in the scanning plane. Moreover, even at a fixed position
the probe sees each point of the AUT from a different angle. As a consequence, the FF of the AUT
cannot be accurately recovered from the measured NF data by employing the uncompensated 
NF-FF transformation. The basic theory of probe compensated NF measurements on a plane as proposed 
in Refs.~\cite{paris1978,joy1978} is based on the application of the Lorentz reciprocity theorem. 
The key relations in the reference system used in the present work are:
\beq
E_\theta(\theta,\phi) = \frac{1}{\Delta} [I_H {E_{\phi}}_V'(\theta,-\phi) - I_V {E_{\phi}}_H'(\theta,-\phi) ] 
\eeq
\beq
E_\phi(\theta,\phi) = \frac{1}{\Delta} [I_H {E_{\theta}}_V'(\theta,-\phi) - I_V {E_{\theta}}_H'(\theta,-\phi) ]
\eeq
where:
\beq
\Delta = {E_{\theta}}_H'(\theta,-\phi) {E_{\phi}}_V'(\theta,-\phi) - 
{E_{\theta}}_V'(\theta,-\phi) {E_{\phi}}_H'(\theta,-\phi)
\eeq
and
\begin{align}
I_{V,H} & =  A \cos\theta \, e^{j \beta d \cos\theta} \times  \nonumber \\
        & \int_{-\infty}^{\infty} \int_{-\infty}^{\infty}  V_{V,H}(x,y) 
\, e^{j \beta x \sin\theta \cos\phi }  
\, e^{j \beta y \sin\theta \sin\phi } {\rm d}x {\rm d}y
\label{eq:ft}
\end{align}
where $A$ is a constant and $\beta$ is the free-space wavenumber. Namely, the 
antenna FF is related to: {\it (i)} the 2D Fourier transforms $I_V$ and $I_H$ 
of the output voltages $V_V$ and $V_H$ of the probe for the
two independent sets of measurements; and 
{\it (ii)} the FF components ${E_{\theta}}_V'$, ${E_{\phi}}_V'$ and 
${E_{\theta}}_H'$, ${E_{\phi}}_H'$ radiated by the 
probe and the rotated probe, respectively, when used as transmitting antennas.
According to Ref.~\cite{yag1984}, the FF components of the electric field, 
${E_{\theta}}_V'$, ${E_{\phi}}_V'$, radiated by an 
open-ended rectangular waveguide (of sizes  
$a'$ and $b'$ along the $x$ and $y$ axis, respectively) excited by a  
TE$_{10}$ mode are:
\beq
{E_{\theta}}_V' = f_\theta(\theta; a',b') \, \sin\phi \frac{e^{-j\beta r} }{r}
\eeq
\beq
{E_{\phi}}_V' = f_\phi(\theta; a',b') \, \cos\phi \frac{e^{-j\beta r} }{r}
\eeq
where the function $f_\phi(\theta; a',b')$ is discussed in 
Ref.~\cite{yag1984}. Similar equations can be found for the 
${E_{\theta}}_H'$, ${E_{\phi}}_H'$ field components. 

%
   \begin{figure}[ht]
   \begin{center}
   \begin{tabular}{c} 
     \includegraphics[height=6cm,angle=0]{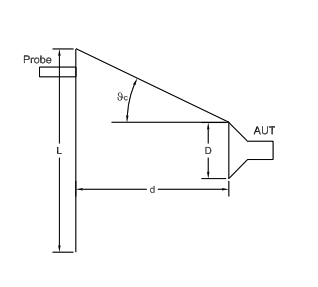}        
   \end{tabular}
   \end{center}
   \caption[example]
{ \label{fig:critang}
Definition of the critical angle, $\theta_c$, and its relation to the 
geometry of the measurement setup. From Fig.~\ref{fig:open2D}, $D=9\,$cm 
and $d=10\,$cm.}
\end{figure}

According to Eq.~(\ref{eq:ft}), in order to obtain the FF pattern all over the 
hemisphere in front of the AUT, the measurement plane should be infinite but, 
of course, this is not possible in a practical setup. The dimension of the plane 
should be such that the field becomes negligible at its edges, thus, minimizing 
the error associated with this truncation. Due to this so-called truncation error, 
the calculated FF using the planar NF data is valid only up to a critical angle  
$\theta_c \simeq 16.7^{\circ}$ outward from the aperture of the AUT 
(see Fig.~\ref{fig:critang}):
\beq
\theta_c = \arctan \left ( \frac{L-D} {2d} \right ) \, .
\eeq
%


\bibliographystyle{spmpsci}      

\bibliography{report}              

\begin{thebibliography}{10}
\providecommand{\url}[1]{{#1}}
\providecommand{\urlprefix}{URL }
\expandafter\ifx\csname urlstyle\endcsname\relax
  \providecommand{\doi}[1]{DOI~\discretionary{}{}{}#1}\else
  \providecommand{\doi}{DOI~\discretionary{}{}{}\begingroup
  \urlstyle{rm}\Url}\fi

\bibitem{born1999}
{Born}, M., {Wolf}, E.: {Principles of Optics}, seventh edn. (1999)

\bibitem{cagigal2004}
{Cagigal}, M.P., {Canales}, V.F., {Oti}, J.E.: {Design of Continuous
  Superresolving Masks for Ground-based Telescopes}.
\newblock Publications of the Astronomical Society of the Pacific \textbf{116},
  965--970 (2004).
\newblock \doi{10.1086/425592}

\bibitem{canales2004}
{Canales}, V.F., {de Juana}, D.M., {Cagigal}, M.P.: {Superresolution in
  compensated telescopes}.
\newblock Optics Letters \textbf{29}, 935--937 (2004).
\newblock \doi{10.1364/OL.29.000935}

\bibitem{cox1982}
{Cox}, I.J., {Sheppard}, C.J.R., {Wilson}, T.: {Reappraisal of arrays of
  concentric annuli as superresolving filters}.
\newblock Journal of the Optical Society of America (1917-1983) \textbf{72}
  (1982)

\bibitem{dejuana2003}
{de Juana}, D.M., {Oti}, J.E., {Canales}, V.F., {Cagigal}, M.P.: {Design of
  superresolving continuous phase filters}.
\newblock Optics Letters \textbf{28}, 607--609 (2003).
\newblock \doi{10.1364/OL.28.000607}

\bibitem{goldsmith1998}
{Goldsmith}, P.F.: {Quasioptical Systems} (1998)

\bibitem{joy1972}
{Joy}, E., {Paris}, D.: {Spatial sampling and filtering in near-field
  measurements}.
\newblock IEEE Transactions on Antennas and Propagation \textbf{20}, 253--261
  (1972).
\newblock \doi{10.1109/TAP.1972.1140193}

\bibitem{joy1978}
{Joy}, E.B., {Leach} Jr., W.M., {Rodrigue}, G.P., {Paris}, D.T.: {Applications
  of probe-compensated near-field measurements}.
\newblock IEEE Transactions on Antennas and Propagation \textbf{26}, 379--389
  (1978).
\newblock \doi{10.1109/TAP.1978.1141856}

\bibitem{kellerer2014}
{Kellerer}, A.: {Beating the diffraction limit in astronomy via quantum
  cloning}.
\newblock Astronomy and Astrophysics \textbf{561}, A118 (2014).
\newblock \doi{10.1051/0004-6361/201322665}

\bibitem{kim2012}
{Kim}, H., {Bryant}, G.W., {Stranick}, S.J.: {Superresolution four-wave mixing
  microscopy}.
\newblock Optics Express \textbf{20}, 6042 (2012).
\newblock \doi{10.1364/OE.20.006042}

\bibitem{liu2008}
{Liu}, L., {Wang}, G.: {Designing superresolution optical pupil filter with
  constrained global optimization algorithm}.
\newblock Optik \textbf{119}, 481--484 (2008)

\bibitem{martinez2002}
{Martinez-Corral}, M., {Caballero}, M.T., {Stelzer}, E.H.K., {Swoger}, J.:
  {Tailoring the axial shape of the point spread function using the Toraldo
  concept}.
\newblock Optics Express \textbf{10}, 98 (2002)

\bibitem{may2004}
{May}, J., {Jennetti}, T.: {Telescope resolution using negative refractive
  index materials}.
\newblock In: UV/Optical/IR Space Telescopes: Innovative Technologies and
  Concepts, \emph{Proc. SPIE}, vol. 5166, p. 220 (2004)

\bibitem{mugnai2003}
{Mugnai}, D., {Ranfagni}, A., {Ruggeri}, R.: {Pupils with super-resolution}.
\newblock Physics Letters A \textbf{311}, 77--81 (2003).
\newblock \doi{10.1016/S0375-9601(03)00445-6}

\bibitem{neil2000}
{Neil}, M.A.A., {Wilson}, T., {Juskaitis}, R.: A wavefront generator for
  complex pupil function synthesis and point spread function engineering.
\newblock Journal of Microscopy \textbf{197}, 219 (2000)

\bibitem{olmi2007}
{Olmi}, L., {Bolli}, P.: {Ray-tracing and physical-optics analysis of the
  aperture efficiency in a radio telescope}.
\newblock Appl. Opt. \textbf{46} (2007)

\bibitem{olmi2017}
{Olmi}, L., {Bolli}, P., {Carbonaro}, L., {Cresci}, L., {Mugnai}, D., {Natale},
  E., {Nesti}, R., {Panella}, D., {Roda}, J., {Zacchiroli}, G.: {Design of
  Super-Resolving Toraldo Pupils for Radio Astronomical Applications}.
\newblock Proc. of the XXXII URSI General Assembly and Scientific Symposium,
  submitted  (2017)

\bibitem{olmi2016}
{Olmi}, L., {Bolli}, P., {Cresci}, L., {Mugnai}, D., {Natale}, E., {Nesti}, R.,
  {Panella}, D., {Stefani}, L.: {Super-resolution with Toraldo pupils: analysis
  with electromagnetic numerical simulations}.
\newblock In: Ground-based and Airborne Telescopes VI, \emph{Proc. SPIE}, vol.
  9906 (2016)

\bibitem{paris1978}
{Paris}, D.T., {Leach} Jr., W.M., {Joy}, E.B.: {Basic theory of
  probe-compensated near-field measurements}.
\newblock IEEE Transactions on Antennas and Propagation \textbf{26}, 373--379
  (1978).
\newblock \doi{10.1109/TAP.1978.1141855}

\bibitem{ranfagni2004}
{Ranfagni}, A., {Mugnai}, D., {Ruggeri}, R.: {Beyond the diffraction limit:
  Super-resolving pupils}.
\newblock Journal of Applied Physics \textbf{95}, 2217--2222 (2004).
\newblock \doi{10.1063/1.1644026}

\bibitem{roy2011}
{Roy}, A., {Ade}, P.A.R., {Bock}, J.J., {Brunt}, C.M., {Chapin}, E.L.,
  {Devlin}, M.J., {Dicker}, S.R., {France}, K., {Gibb}, A.G., {Griffin}, M.,
  {Gundersen}, J.O., {Halpern}, M., {Hargrave}, P.C., {Hughes}, D.H., {Klein},
  J., {Marsden}, G., {Martin}, P.G., {Mauskopf}, P., {Netterfield}, C.B.,
  {Olmi}, L., {Patanchon}, G., {Rex}, M., {Scott}, D., {Semisch}, C., {Truch},
  M.D.P., {Tucker}, C., {Tucker}, G.S., {Viero}, M.P., {Wiebe}, D.V.:
  {Deconvolution of Images from BLAST 2005: Insight into the K3-50 and IC 5146
  Star-forming Regions}.
\newblock Astrophysical Journal \textbf{730}, 142 (2011).
\newblock \doi{10.1088/0004-637X/730/2/142}

\bibitem{silver}
{Silver}, S.: {Microwave Antenna theory and Design}.
\newblock McGraw-Hill Company (1949)

\bibitem{toraldo1952a}
{Toraldo di Francia}, G.: Il Nuovo Cimento (Suppl.) \textbf{9}, 426 (1952)

\bibitem{toraldo1952b}
{Toraldo di Francia}, G.: Atti Fond. Giorgio Ronchi \textbf{7}, 366 (1952)

\bibitem{yag1984}
{Yaghjian}, A.D.: {Approximate formulas for the far field and gain of
  open-ended rectangular waveguide}.
\newblock IEEE Transactions on Antennas and Propagation \textbf{32}, 378--384
  (1984).
\newblock \doi{10.1109/TAP.1984.1143332}

\bibitem{zhang2008}
{Zhang}, X., {Liu}, Z.: {Superlenses to overcome the diffraction limit}.
\newblock Nature Materials \textbf{7}, 435--441 (2008).
\newblock \doi{10.1038/nmat2141}

\end{thebibliography}

\end{document}